\documentclass[a4paper,fleqn,usenatbib]{mnras}

\usepackage[T1]{fontenc}
\usepackage{ae,aecompl}

\usepackage{graphicx}	
\usepackage{amsmath}	
\usepackage{amssymb}	

\usepackage{caption}
\usepackage{cuted}

\usepackage{lscape}

\usepackage{changepage}
\usepackage{mathtools}
\usepackage{booktabs}
\usepackage{adjustbox}

\DeclareMathOperator{\sgn}{sgn}

\title[A local model for Spherical Flows]{A Local Model for the Spherical Collapse/Expansion Problem}

\author[Elliot M. Lynch and Guillaume Laibe]{
Elliot M. Lynch,$^{1}$\thanks{E-mail: elliot.lynch@ens-lyon.fr}
Guillaume Laibe,$^{1}$
\\
$^{1}$Univ Lyon, Univ Lyon1, Ens de Lyon, CNRS, Centre de Recherche Astrophysique de Lyon UMR5574, F-69230, Saint-Genis,-Laval, France\\
}

\date{Accepted XXX. Received YYY; in original form ZZZ}

\pubyear{2019}

\begin{document}
\label{firstpage}
\pagerange{\pageref{firstpage}--\pageref{lastpage}}
\maketitle

\begin{abstract}
Spherical flows are a classic problem in astrophysics which are typically studied from a global perspective. However, much like with accretion discs, there are likely many instabilities and small scale phenomena which would be easier to study from a local perspective. For this purpose, we develop a local model for a spherically contracting/expanding gas cloud, in the spirit of the shearing box, $\beta$-plane and expanding box models which have had extensive use in studies of accretion discs, planets and stellar winds respectively. The local model consists of a, spatially homogeneous, periodic box with a time varying aspect ratio, along with a scale factor (analogous to that in FRW/Newtonian cosmology) relating the box coordinates to the physical coordinates of the global problem. We derive a number of symmetries and conservation laws exhibited by the local model. Some of these reflect symmetries of the periodic box, modified by the time dependant geometry, while others are local analogues for symmetries of the global problem. The energy, density and vorticity in the box also generically increase(/decrease) as a consequence of the collapse(/expansion). We derive a number of nonlinear solutions, including a local analogue of uniform density zonal flows, which grow as a consequence of angular momentum conservation. Our model is closely related to the accelerated expanding box model of Tenerani \& Velli and is an extension of the isotropic model considered by Robertson \& Goldreich.
\end{abstract}




\section{Introduction}

Spherically expanding/collapsing flows are a classic problem are a classic problem in astrophysics. Such flows include the classic Bondi spherical accretion problem \citet{Bondi52}; the vonNeumann-Sedov-Taylor solution for a spherical shock wave, used as a model of supernova explosions \citet{Sedov46,Taylor46,Taylor50,Bethe58}; solar and stellar winds \citep{Velli92,Grappin93,Grappin96,Tenerani17,Shi20,Huang22} and the spherical collapse problem \citep{Larson69,Penston69,Shu77,Hunter77,Foster93} important for star and planet formation. While in many of these problems the spherically symmetric flow provides a useful approximation to the leading order dynamics, one expects there to be local departure from spherical symmetry that could be important in many applications. This is particularly true of the collapse problem as rotational flows are expected to grow during the collapse due to conservation of angular momentum (see also \citet{Velli92,Grappin93,Grappin96,Tenerani17,Shi20,Huang22} for an application where local flows are important in an expanding flow).

One can study local departures from the axisymmetric flows in the global picture, such as by performing full 3D simulations of the expansion/collapse. However it can be prohibitively computationally expensive to resolve both the scale of interest, while simulating the entire spherical expansion/collapse. Instead one can turn to local models, which are commonly used in astrophysics and planetary sciences to tackle such problems. Commonly used local models include the shearing box model \citep{Goldreich65,Hawley95,Latter17}, used in the study of accretion discs; the $\beta$-plane model \citet{Rossby39}, used to study rotating planets; and, of particular relevance to our study here, the Expanding Box Model \citep[EBM][]{Velli92,Grappin93,Grappin96} and Accelerated Expanding Box \citep[AEB][]{Tenerani17}, used to study stellar winds and the FRW like model of \citet{Robertson12} used to study turbulence in cosmological (i.e. homogeneous and isotropic) collapses.  Such local models have had a number of notable successes in astrophysics, the most famous of which is the (re-)discovery of MRI by \citet{Hawley95}.

Of the existing local model the most relevant to our work is the EBM model for stellar winds, developed by \citet{Velli92,Grappin93,Grappin96}. This model follows a local box in a (supersonic), uniformly, expanding magnetohydodynamic (MHD) flow to study local instabilities and waves in the outer regions of solar and stellar winds. In this paper we are interested in deriving an expanding box like model valid for both radially and temporally varying spherical flows. The generalisation of the expanding box model to radially varying flows was done in \citet{Tenerani17}. The generalisation to background flows which are also time dependant, motivated by the stellar formation problem, results in a model close to the accelerated expanding box (although our treatment of pressure will be closer to the distorted shearing box models of \citet{Ogilvie13a,Ogilvie14}). We shall focus, in this paper, on the hydrodynamic case as it posses a number of important features that are worth understanding before generalising to MHD. 

In Section \ref{derivations} we present the derivation of our local model. Sections \ref{symmetries} and \ref{conservation laws} derives symmetries and conservation laws of the local model. Section \ref{nonlinear solutions} presents some nonlinear solutions to the local model - and discuss how these relate to the global problem. In Section \ref{linear perturbations} we derive the linear theory of our local model. We discuss possible extension of our model in Section \ref{discussion}. We present our conclusions in Section \ref{conclusion} and additional mathematical details (including alternative formulations which maybe more convenient for implementation in hydrocodes) are presented in the appendices.

\section{Derivation} \label{derivations}

\subsection{Global geometry}

\begin{figure}
\includegraphics[trim=0 450 0 10, clip, width=\linewidth]{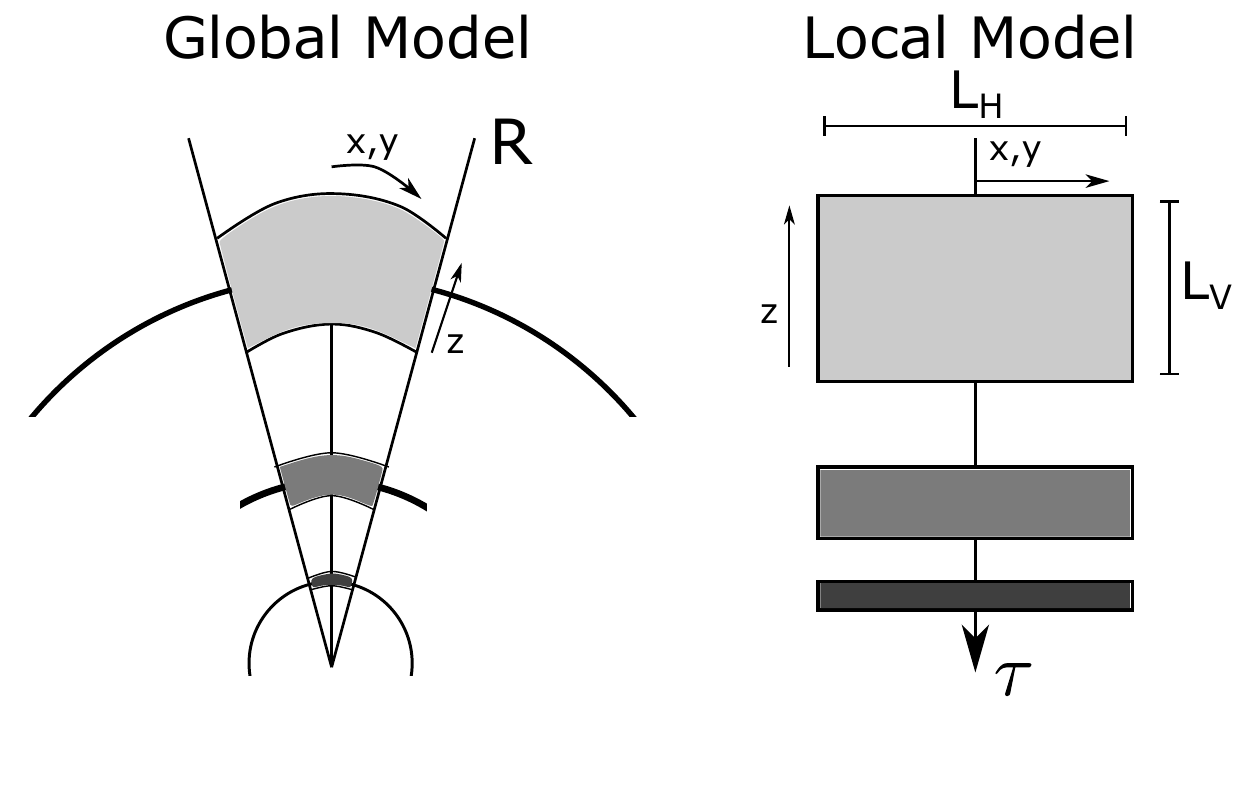}
\caption{Geometry of the domain. Globally (left) the domain is bounded by radial shells which can approach or recede from each other depending on the gradients in the background velocity. Points within the domain move in the radial direction due to the spherically symmetric background flow. The local model (right) is a rectangular domain where the horizontal coordinates are equivalent to the latitude/longitude on the sphere and the vertical direction moves between spherical shells comoving with the background flow. The aspect ratio of this local box changes as the distance between the spherical shells varies.}
\label{collapse geom}
\end{figure}

To derive a local model for spherical collapse/expansion consider a local neighbourhood of a point, $o$, located on the equator of a sphere of radius $R$. The line element of the usual spherical polar coordinate system is

\begin{equation}
 \mathrm{d} s^2 =  \mathrm{d} R^2 + R^2 ( \mathrm{d} \theta^2 + \sin^2 \theta \,  \mathrm{d} \phi^2) .
\end{equation}
We are interested in describing the local dynamics near to $p$ occurring on a horizontal lengthscale $L_{\rm H} \ll R$ (See Figure \ref{collapse geom}, which show the relationship between the global and local geometries). Without loss of generality, we can locate our local model on the equator of the sphere ($\theta = \pi/2$) meaning we can approximate the line element by
\begin{equation}
  \mathrm{d} s^2 =  \mathrm{d} R^2 + R^2 ( \mathrm{d} \theta^2 +  \mathrm{d} \phi^2) + \mathcal{O}( (L_{\rm H} / R)^2 d \phi^2 ) ,
\end{equation}
which results in metric tensor components,

\begin{equation}
g_{R R} = 1 , \quad g_{\theta \theta} = g_{\phi \phi} = R^2 , 
\end{equation}
and inverse metric tensor components
\begin{equation}
g^{R R} = 1 , \quad g^{\theta \theta} = g^{\phi \phi} = R^{-2} , 
\end{equation}
with all other components zero. The Christoffel Symbols components, for this coordinate system, are

\begin{align}
\begin{split}
 \Gamma_{\theta \theta}^{R} &= \Gamma_{\phi \phi}^{R} = - R , \\
 \Gamma_{\theta R}^{\theta} =  \Gamma_{R \theta}^{\theta} &= \Gamma_{\phi R}^{\phi} =  \Gamma_{R \phi}^{\phi}  = R^{-1} ,
\end{split}
\end{align}
with all others vanishing. The fluid equations in this coordinate system are

\begin{align}
D u^{\theta} + \frac{2}{R} u^{R} u^{\theta}  &= - R^{-2} \left( \partial_{\theta} \Phi + \frac{1}{\rho} \partial_{\theta} p \right) , \\
D u^{\phi} + \frac{2}{R} u^{R} u^{\phi}  &= - R^{-2} \left( \partial_{\phi} \Phi + \frac{1}{\rho} \partial_{\phi} p \right) , \\
D u^{R} - R u^{\phi} u^{\phi} - R u^{\theta} u^{\theta}  &= - \left( \partial_{R} \Phi + \frac{1}{\rho} \partial_{R} p \right) ,  \\
D \rho &= - \rho R^{-2} \partial_{i} (R^2 u^i) ,
\end{align}
where the Lagrangian derivative is

\begin{equation}
D = \partial_t + u^{i} \partial_{i} .
\end{equation}
Note, we have listed the $R$ component of the momentum equation last as it will become the $z$ momentum equation in the local coordinate system. To close this system of equations we must supplement them with an equation of state determining $p$, which we assume is barotropic,

\begin{equation}
p = p (\rho) .
\end{equation}

\subsection{Spherical Collapse/Expansion}

For the background fluid flow we wish to consider a spherically symmetric expanding/contracting fluid in a (potentially time dependant) central potential $\Phi = \Phi(t,R)$. Consider a spherically symmetric fluid in this potential with density $\rho_0 = \rho_0 (R, t)$ and purely radial velocity field $U^{i} = U (R, t) \hat{e}_{R}^{i}$. The density of the fluid then evolves according to the continuity equation,

\begin{equation}
D_{0} \rho_0 = - \rho_{0} R^{-2} \partial_{R} (R^2 U ) ,
\end{equation}
where the Lagrangian derivative of the background flow $D_0$ is given by,

\begin{equation}
D_{0} = \partial_{t} + U \partial_{R} .
\end{equation}
The radial momentum equation for the background flow determines how $U(R, t)$ evolves,

\begin{equation}
D_{0} U = - \partial_{R} \Phi - \frac{1}{\rho_0} \partial_{R} p_0 ,
\label{background eom}
\end{equation}
where $p_0 = p (\rho_0)$ is the fluid pressure. Spherical symmetry ensures that the $\theta$ and $\phi$ components of the momentum equation are satisfied.

This background flow naturally sets a characteristic timescale, $t_{\rm bg} \sim R/|U|$, which is the timescale over which the background flow evolves (in the AEB model of \citet{Tenerani17} their expansion timescale $\tau_{e}$ is equal to our $t_{\rm bg}$). In the collapse case we also have $t_{\rm collapse}$ (typically $\sim t_{\rm bg}$), which is the timescale for the completion of the collapse (e.g. the free-fall time or $t_{\rm collapse} \sim t_{\rm bg} \sim \frac{R (0)}{c_s}$ in the isothermal sphere model of \citet{Shu77}). In general one expects $t_{\rm bg}$ to evolve with the background flow, typically getting longer in expanding flows and shorter for collapses.

\subsection{Deriving the Local Model}

The local model we shall derive is similar to the AEB \citep{Tenerani17}, however we allow for a time dependant background flow and our treatment of pressure is closer to the distorted shearing box models of \citet{Ogilvie13a,Ogilvie14}. We are also interested in what happens when such a model is run in reverse as a model of a spherical collapse.

Consider local, nonlinear, perturbations to the background, spherically symmetric, flow. We now write the fluid motion as the sum of the background spherical collapse/expansion and a relative velocity $v^{i}$:

\begin{equation}
u^{i} = U \hat{e}^{i}_{R} + v^{i} .
\label{background local split}
\end{equation}
The fluid dynamical equations for the relative velocity are, without approximation,

\begin{align}
D v^{\theta} + \frac{2}{R} (U + v^{R}) v^{\theta}  &= \frac{1}{\rho_0 R^2} \partial_{\theta} p_0 - \frac{1}{\rho R^2} \partial_{\theta} p  , \\
D v^{\phi} + \frac{2}{R} (U + v^{R}) v^{\phi}  &= \frac{1}{\rho_0 R^2} \partial_{\phi} p_0 - \frac{1}{\rho R^2} \partial_{\phi} p  , \\
D v^{R} +  v^{R} \partial_{R} U  - R v^{\phi} v^{\phi} - R u^{\theta} u^{\theta}  &= \frac{1}{\rho_0} \partial_{R} p_0 - \frac{1}{\rho} \partial_{R} p , \\
D \rho &= - \rho [\Delta + R^{-2} \partial_{i} (R^2 v^{i})]  ,
\end{align}
where the Lagrangian derivative is

\begin{equation}
D = \partial_t + U \partial_{R} + v^i \partial_{i} ,
\end{equation}
and we have introduced the velocity divergence of the background flow,

\begin{equation}
\Delta = R^{-2} \partial_{R} (R^2 U) .
\end{equation}

As the fluid is barotropic, we can introduce the pseudo-enthalpy $h (\rho) =  \int p(\rho) d \rho^{-1}$, and the momentum equations can now be written as

\begin{align}
D v^{\theta} + \frac{2}{R} (U + v^{R}) v^{\theta}  &=  - \frac{1}{R^2} \partial_{\theta} h  , \\
D v^{\phi} + \frac{2}{R} (U + v^{R}) v^{\phi}  &= - \frac{1}{R^2} \partial_{\phi} h  , \\
D v^{R} +  v^{R} \partial_{R} U  - R v^{\phi} u^{\phi} - R u^{\theta} u^{\theta}  &= - \partial_{R} h - \partial_{R} h_0 ,
\end{align}
where $h_0 = h (\rho_0)$.

We are interested in describing local, nonlinear, perturbations occurring on a horizontal lengthscale $L_{\rm H} \ll R$ and vertical lengthscale $L_{\rm V} \ll R$, where the nonlinear curvature terms $(\frac{2}{\mathrm{R}} v^{R} v^{\phi}, \frac{2}{\mathrm{R}} v^{R} v^{\theta} , R (v^{\phi} v^{\phi} + v^{\theta} v^{\theta}))$ can be neglected at leading order. For a hypersonic collapse/expansion these terms are subdominant relative to the global curvature terms $\frac{2 U}{R}$ and the vertical advection of the background flow, $v^{R} U_{0 R}$ (assuming $U_{R 0} \sim U/R$). This is the regime considered in the expanding box model of \citet{Velli92,Grappin93,Grappin96} \footnote{There is a subtlety in the expanding box model as the uniform expansion means that $U_{R 0} = 0$ so that, strictly, $R (v^{\phi} v^{\phi} + v^{\theta} v^{\theta})$ cannot, in general, be neglected. In the expanding box application one is saved by the fact that $v^{\phi}$, $v^{\theta}$ tend to get small with the expansion as a consequence of angular momentum conservation meaning if they are small enough initially the nonlinear curvature terms should remain small throughout the expansion. This is not the case if one were to run the expanding box in reverse to model a uniform collapse and thus extra care must be taken in this instance.}. It also results in a similar asymptotic ordering scheme to that used to derive the shearing box (in particular the ``distorted'' eccentric and warped variants, \citet{Ogilvie13a,Ogilvie14}). However, typical collapse profile \citep[such as][]{Shu77} start out subsonic and transition to a hypersonic flow at late times and one would ideally like our local model to be able to handle this situation. In the highly subsonic regime the local flows primarily consist of sound and vortical waves and the local model is (at leading order) a Cartesian, periodic, box with an anisotropic sound speed. The main influence of the background flow on the model, in this regime, is the slow variation of the horizontal and vertical sound speed with time. Thus the nonlinear curvature term is subdominant to the terms involving pressure gradients.

In general then, in order that we can neglect the nonlinear curvature terms we require,

\begin{align}
R v^{R} v^{\phi} &\ll \max (v^{\phi} R U , (L_{\rm H}/R)^{-1} c_s^2) , \label{phi curv cond}\\
R v^{R} v^{\theta} &\ll \max (v^{\theta} R U , (L_{\rm H}/R)^{-1} c_s^2) , \\
R (v^{\phi} v^{\phi} + v^{\theta} v^{\theta} ) &\ll \max (v^{R} R U_{R 0} , (L_{\rm V}/R)^{-1} c_s^2) . \label{r curv cond}
\end{align}
To ensure that these terms are subdominant to either the terms involving the pressure gradient or the terms involving the background flow.

This requirement that the nonlinear curvature terms be negligible tends to introduce a timescale, $t_{\rm curv}$,  over which the local approximation is valid and timescales longer than $t_{\rm curv}$ the neglected terms cause a significant departure from the predictions of the local model. This is particularly true for the collapse case $(U < 0)$ as the nonlinear curvature terms are expected to grow with time, eventually violating Conditions \ref{phi curv cond}-\ref{r curv cond}. This just reflects the expectation that (hydrodynamic) collapses tend to be halted when they achieve sufficient rotational support. Thing are easier for spherical expansions as the neglected terms tend to get smaller with the expansion. If $t_{\rm curv} \gtrsim t_{\rm collapse} \sim t_{\rm bg}$ then we can successfully employ the local model over most of the collapse. However when $t_{\rm curv} < t_{\rm collapse} \sim t_{\rm bg}$ our model breaks down before the collapse completes. While this is a limitation of the local model it's worth noting that physically if $t_{\rm curv} < t_{\rm collapse}$ this means the background spherical collapse is also not valid on timescales longer than $t_{\rm curv}$ as local/nonspherical perturbations have grown sufficiently nonlinear to modify the background collapse.

In addition to the nonlinear curvature terms we must also consider how to deal with the background pressure gradient, $-\partial_R h_0$. In the small box limit that we are considering, this term is always subdominant relative to the main pressure gradient term $-\partial_R h$. In the hypersonic limit this term is also subdominant to the background advection terms and it is evident that one can neglect this term. The hypersonic limit (neglecting the $-\partial_R h_0$ term) is a consistent asymptotic limit of the background flow similar to the `small' compressible shearing box as described in \citet{Latter17}. It is less obvious what should be done in the subsonic limit as here the advection terms are also subdominant. \citet{Tenerani17} have argued that both the background pressure gradient and background advection terms should be maintained as they are of the same order. However, it has been found that keeping the pressure gradient terms leads to spurious instabilities such as unstable sound waves\footnote{It's unclear whether the gradients in the background magnetic field in the model of \citet{Tenerani17} will lead to similar spurious instabilities in the magnetosonic waves.} \citep[See][for a discussion in the shearing box context]{Latter17}. This may be a consequence of the fact the basic state of the local model including the background pressure gradient is incompatible with the imposition of periodic boundaries that are typically required for numerical applications. The advection terms, on the other hand, are a relatively benign subdominant term in the subsonic limit. In fact on can show that the effects of the advection in the subsonic limit are equivalent to their effects on short-wavelength waves in the hypersonic limit. Therefore the correct approach appears to be to neglect the background pressure gradient, but maintain the background advection terms, provided that the box is small enough.

Neglecting the nonlinear curvature terms, and background pressure gradient, the equations simplify to

\begin{align}
D v^{\theta} + \frac{2}{R} U v^{\theta}  &=  - \frac{1}{R^2} \partial_{\theta} h  , \\
D v^{\phi} + \frac{2}{R} U v^{\phi}  &= - \frac{1}{R^2} \partial_{\phi} h   , \\
D v^{R} + v^{R} \partial_{R} U  &= - \partial_{R} h  , \\
D \rho &= - \rho [\Delta +  \partial_{i} v^{i}]  .
\end{align}

Consider a reference point following the spherical flow, starting at an initial radius $R_0$, and angular coordinates $\theta = \pi/2$, $\phi = 0$ at $t = 0$. Let $\mathcal{R} (t)$ be the solution of $d \mathcal{R} (t)/d t = U(\mathcal{R} (t), t)$ subject to the initial condition $\mathcal{R} (0) = R_0$. Then the coordinates of the reference point are $(\mathcal{R} (t),\theta, \phi) = (R_0,\pi/2,0)$. We now consider a local neighbourhood of the reference point, 

\begin{equation}
 \theta = \pi/2 + x , \quad \phi = y , \quad R = \mathcal{R} (t) + z , \quad t = \tau ,
\end{equation}
where $x$, $y$ are $\mathcal{O}(L_{\rm H}/\mathcal{R})$ and $z$ is $\mathcal{O}(L_{\rm V}/\mathcal{R})$ in our units. Thus our coordinate system consists of a (small) rectangular domain on the equator of a sphere of radius $\mathcal{R}$, with $x, y$ being Cartesian coordinate describing the location of the point on this sphere and $z$ denoting the height above the reference sphere along lines of radius. The radius of the reference sphere is then free to change with time - accounting for the spherical collapse/expansion in the global system.

Because of it's appearance in the Lagrangian derivative we must expand the background velocity vertically as follows:

\begin{equation}
U = U_0 (\tau) + U_{R 0}  (\tau) z .
\end{equation}
All other geometrical/background quantities are evaluated at the reference point and are thus functions of time only.

Because of the time dependence of the coordinate system the old and new time derivatives are related by
\begin{equation}
 \partial_t + U_0 \partial_{R} = \partial_{\tau} ,
\end{equation}
while the Lagrangian time derivative is 

\begin{equation}
 D = \partial_t + U_{R 0} z \partial_{z} + v^{i} \partial_i .
\end{equation}
The resulting fluid equations in this coordinate system are then

\begin{align}
D v^{x} + \frac{2 U_0}{\mathcal{R}} v^{x}  &=  - \frac{1}{\rho \mathcal{R}^2} \partial_{x} p , \\
D v^{y} + \frac{2 U_0}{ \mathcal{R}} v^{y}  &= - \frac{1}{\rho \mathcal{R}^2} \partial_{y} p  , \\
D v^{z} + v^{z} U_{R 0}  &= - \frac{1}{\rho} \partial_{z} p  ,
\end{align}
where $p = p(\rho)$ is given by the barotropic equation of state. 

One drawback of the above coordinate system is that, excepting for the case of a uniformly contracting/expanding fluid, there is an explicit dependence on the vertical coordinate through the Lagrangian time derivative. This makes it harder to setup boundary conditions in the vertical direction. Similar to local models of distorted discs \citep[e.g.][]{Ogilvie14}, we can rectify this deficiency by adopting a Lagrangian or Stretched vertical coordinate $\tilde{z} = z/L_z (\tau)$. $\tilde{z}$ is a Lagrangian coordinate with respect to the background flow, i.e. $D_0 \tilde{z} = 0$, where $D_0 = \partial_{\tau} + U_{R 0} z \partial_z$. $L_z$ is a characteristic vertical lengthscale which encompasses the vertical stretching/compression of the fluid flow due to radial variations of the background velocity $U$. $L_z$ evolves according to 

\begin{equation}
 \frac{d L_z}{d \tau} = U_{R 0} L_z ,
 \label{Lz evolution}
\end{equation}
In principle one can rescale the vertical coordinate such that $L_z (0) = R_0$. However taking $L_z (0) = L_{z 0}$, which need not equal $R_0$, allows for the exploration of flows with different horizontal and vertical lengthscales.

The vertical partial derives are related by $\partial_{z} = L_z^{-1} \partial_{\tilde{z}}$ and the vertical velocities are related by $v^{\tilde{z}} = L_{z}^{-1} v^{z}$. Upon adopting the stretched vertical coordinates the Lagrangian time derivative transforms to a spatially homogeneous form,

\begin{equation}
D = \partial_{\tau} + v^{x} \partial_x + v^{y} \partial_y + v^{\tilde{z}} \partial_{\tilde{z}} .
\end{equation}
The Jacobian determinant of the new coordinate system is $J = L_z \mathcal{R}^{2}$ and the coordinate system has the following line element

\begin{equation}
d s^2 = \mathcal{R}^2 (\tau) [d x^{2}  + d y^{2} ] + L_z^2 (\tau) d \tilde{z}^{2} .
\end{equation}
This choice of vertical coordinate puts the vertical and horizontal coordinates on equal footing. Both are dimensionless variables with an associated lengthscale ($L_z$ and $\mathcal{R}$ respectively). Using the relationship between $\mathcal{R}$ and $U_0$, along with Equation \ref{Lz evolution} we can write $\Delta$ in terms of the time derivative of the Jacobian,

\begin{equation}
 \Delta = \frac{2 U_0}{\mathcal{R}} + U_{R 0} = \frac{2}{\mathcal{R}} \frac{d \mathcal{R}}{d \tau} + \frac{1}{L_z} \frac{d L_z}{d \tau} = \frac{1}{J} \frac{d J}{d \tau} .
\end{equation}

In the stretched coordinate system the momentum and continuity equations of the local model are

\begin{align}
D v^{x} + \frac{2 U_0}{\mathcal{R}} v^{x} &= - \frac{1}{\rho \mathcal{R}^{2}} \partial_{x} p , \label{col box x} \\
D v^{y} + \frac{2 U_0}{\mathcal{R}} v^{y} &= - \frac{1}{\rho \mathcal{R}^{2}} \partial_{y} p , \\
D v^{\tilde{z}} + 2 v^{\tilde{z}} U_{R 0} &= - \frac{1}{\rho L_z^2} \partial_{\tilde{z}} p , \\ 
D \rho &= - \rho \left[ \Delta + \partial_x v^{x} + \partial_y v^{y} + \partial_{\tilde{z}} v^{\tilde{z}} \right] . \label{col box rho}
\end{align}
Equations \ref{col box x}-\ref{col box rho} form our local model. As explored in subsequent sections, one unusual property of this model, that can be deduced from the form of these equations, is the differing effective sounds speeds for sound waves propagating in the vertical and horizontal directions when $L_{z} \ne \mathcal{R}$.
 
Alternatively the momentum equation can be written in terms of the covariant velocity, $v_i$, which are in many ways simpler,

\begin{align}
D v_{x} &= - \frac{1}{\rho} \partial_{x} p ,\\
D v_{y} &= - \frac{1}{\rho} \partial_{y} p , \\
D v_{\tilde{z}} &= - \frac{1}{\rho} \partial_{\tilde{z}} p .
\end{align}
Notably this implies conservation of the generalised momenta, $\rho v_i$.

With this choice of coordinate systems we can now use reflective/closed boundaries (for no mass/momentum flux with neighbouring radial shells) or periodic boundaries in the vertical direction. In the horizontal direction periodic boundaries are the physically meaningful boundary conditions as these allow use to consider the behaviour of high-m perturbations to spherical flows in a local model. The possibility to use periodic boundaries significantly simplify numerical implementation of the model.

A more general choice of boundary condition which will allow for the study of a much wider class of flows is to adopt shear-periodic boundary conditions analogous to those seen in shearing box models, provided the shear across the box is not too large such that the curvature terms become important. Using shear periodic boundary conditions likely allows for the modelling of a weakly rotating collapse/expansion within the local framework. A proper exploration of the shear-periodic local model and it's correspondence with a weakly rotating spherical flows is beyond the scope of this paper and will be left for future work.

The local model can be derived from the following Lagrangian density

\begin{equation}
 \mathcal{L} =  \rho_0 \left[\frac{1}{2} \mathcal{R}^2 (v^x v^x + v^y v^y) + \frac{1}{2} L_z^2 v^{\tilde{z}} v^{\tilde{z}} - \varepsilon \right] ,
\label{nonlinear Lagrangian}
\end{equation}
where $\varepsilon$ is the specific internal energy.

We can also reformulate the local model in terms of a FRW like metric:

\begin{equation}
d s^2 = a^2 (\tau) [d x^{2}  + d y^{2} + b^2 (\tau^{\prime}) d \tilde{z}^{2} ]  ,
\end{equation}
where $a$ is the scale factor and $b$ the box aspect ratio. One has the choice to absorb the dimensions into the scale factor or the coordinate system. $a$ and $b$ are related to $\mathcal{R}$ and $L_z$ through

\begin{equation}
 a = \mathcal{R}, \quad b = L_{z}/\mathcal{R} .
\end{equation}
This leads to the following fluid equations

\begin{align}
D v^{x} + 2 H v^{x} &= - \frac{a^{-2}}{\rho} \partial_{x} p ,\\
D v^{y} + 2 H v^{y} &= - \frac{a^{-2}}{\rho} \partial_{y} p , \\
D v^{\tilde{z}} + 2 H v^{\tilde{z}} + 2 \frac{\dot{b}}{b} v^{\tilde{z}} &= - \frac{a^{-2} b^{-2}}{\rho} \partial_{\tilde{z}} p , \\ 
D \rho &= - \rho \left[ 3 H + \frac{\dot{b}}{b} + \partial_x v^{x} + \partial_y v^{y} + \partial_{\tilde{z}} v^{\tilde{z}} \right] .
\end{align}
where we have defined the Hubble parameter $H = \dot{a}/a$. For a constant aspect ratio box, $\dot{b}=0$, these equations are equivalent to \citet{Robertson12}. 

Various alternative formulations of the local model, of potential use for numerical implementations, are given in Appendix \ref{alternative formalisms}. Of note is the time dependant background terms in the continuity equation, and in front of the pressure gradients, can be absorbed into a time dependant, anisotropic, effective sound speed. 

\subsection{Symmetries of the local model}  \label{symmetries}

Ideal gas dynamics has a number of important symmetries, such as Galilean transforms, rotations, and lengthscale, timescale and mass rescaling. The global problem similarly has many symmetries - in particular global rotational symmetries along with the choice of rotating frame. One would also expect that many of these symmetries should be reflected in the local model. In this section we shall explore which of these symmetries are carried over/modified in the local model.

Fluid dynamics in periodic, Cartesian boxes are invariant under Galilean transforms. In the local model one expect such Galilean transforms will be modified by the time dependant geometry. Consider a modified Galilean transform where the new position and timescale are related to the old ones by

\begin{equation}
  \mathbf{x}^{\prime} = \mathbf{x} - \int \mathbf{v}_g (\tau) d \tau , \quad \tau^{\prime} = \tau .
\end{equation}
where $\mathbf{v}_g (t)$ is the new frame velocity which is uniform across the box, but can undergo acceleration. The velocity transforms according to

\begin{equation}
 \mathbf{v}^{\prime} = \mathbf{v} - \mathbf{v}_g .
\end{equation}
This results in partial derivatives which transform according to

\begin{equation}
 \partial_{\tau} = \partial_\tau^{\prime} - v^{i}_g \partial_{i} , \quad \partial_{x} = \partial_{x}^{\prime} .
\end{equation}
The Lagrangian time derivative is unchanged by this transform with $D = \partial^{\prime}_t + v^{i \prime} \partial_{i}^{\prime}$, similarly the divergence of the relative velocity is left unchanged ($\partial_{i} v^i = \partial_{i}^{\prime} v^{i \prime}$). This means the continuity equation is unchanged by the transform. In order for the momentum equations to remain unchanged by this transform we require that the frame velocity, $\mathbf{v}_g$, satisfy

\begin{align}
 \dot{v}_g^{x} + \frac{2 U_0}{\mathcal{R}} v_g^{x}  &= 0 , \\ 
 \dot{v}_g^{x} + \frac{2 U_0}{\mathcal{R}} v_g^{y} &= 0 , \\ 
 \dot{v}_g^{\tilde{z}} + 2 U_{R 0} v_g^{\tilde{z}} &= 0 .
\end{align}
We see that, during a collapse ($U_0 < 0$), horizontal frame translations accelerate with time in order to conserve angular momentum in the global frame. These horizontal translations in the local frames thus correspond to a (slow) global rotation of the reference sphere. The vertical frame translation correspond to a acceleration of the reference sphere relative to the background flow - i.e. the reference sphere falls/rises at a slightly faster/slower rate than the background flow.

Ideal gas dynamics in Cartesian geometry exhibit a similarity transform where the dynamics are invariant under rescaling of the length/timescale. The local model exhibits a similar similarity transform, however here we must be careful to also perform an appropriate rescaling of the background flow. Consider a rescaling of space, time and fluid entropy by constant factors such that the length and timescale transform like

\begin{equation}
 \mathbf{x} \mapsto \lambda  \mathbf{x}  , \quad \tau \mapsto \mu \tau ,
\end{equation}
while (assuming a perfect gas) the pressure transforms like

\begin{equation}
 p \mapsto \kappa p .
\end{equation}
Under this rescaling, the partial derivatives transform according to

\begin{equation}
 \partial_t \mapsto \mu^{-1} \partial_t, \quad \partial_i \mapsto \lambda^{-1} \partial_i ,
\end{equation}
and the velocity transforms according to

\begin{equation}
 \mathbf{v} \mapsto \lambda \mu^{-1} \mathbf{v} .
\end{equation}
This results in the Lagrangian time derivative transforming like $D \mapsto \mu^{-1} D$. This transformation works provided that the background flow is also transformed like

\begin{equation}
 \mathcal{R} \mapsto (\kappa/\lambda)^{-1/2} \mathcal{R}, \quad L_{z} \mapsto (\kappa/\lambda)^{-1/2} L_{z} , 
\end{equation}
\begin{equation}
 U_0 \mapsto (\kappa/\lambda)^{-1/2} \mu^{-1}  U_0, \quad  U_{R 0} \mapsto \mu^{-1}  U_{R 0}   .
\end{equation}
This also results in $\Delta \mapsto \mu^{-1} \Delta$. This allows us to relate the local flows of two different, but homologous, spherical flows by a rescaling of the lengthscale, timescale and entropy of the local flow.

Independently of this similarity transform one can also rescale the vertical lengthscale of the local model - corresponding to changing the aspect ratio of the box. Consider a rescaling of the vertical lengthscale, $L_z$, by a constant factor $r$ with $ L_z \mapsto r^{-1} L_z$. If we simultaneously rescale the stretched vertical coordinate by

\begin{equation}
 \tilde{z} \mapsto r \tilde{z} ,
\end{equation}
then as, physical vertical coordinate $z = \tilde{z} L_{z}$ is left unchanged by this transform, the dynamics are unchanged by this rescaling. i.e. if we rescale $L_z$ by a constant factor then the local dynamics are left unchanged if we simultaneously rescale the vertical extent of the box. Under this change of aspect ratio the vertical partial derivative transforms as $\partial_{\tilde{z}} \mapsto r^{-1} \partial_{\tilde{z}}$, while the vertical velocity transforms according to

\begin{equation}
 v^{\tilde{z}} \mapsto r v^{\tilde{z}} ,
\end{equation} 
meaning $D$ is unchanged by this change of aspect ratio. One consequence of this rescaling, however, is the velocity field in the new coordinate system may posses shear where the original contained none. This will be particularly important to the diagonally propagating sound waves discussed in subsequent section.

Finally the local model is invariant under mass rescaling ($\rho \mapsto \lambda \rho$, $p \mapsto \lambda p$), assuming a perfect gas, and horizontal rotations. Unlike fluid dynamics in a periodic, Cartesian, box, the local model is not symmetric to rotations in the vertical direction unless the aspect ratio of the box is fixed ($L_z \propto \mathcal{R}$). This is a consequence of the different effective soundspeed in the vertical and horizontal direction, along with the differing contributions from the background velocity.

\subsection{Conservation Laws}  \label{conservation laws}

The conservative form of the continuity equation in the local model is
\begin{equation}
 \partial_{\tau} (J \rho) + \partial_{x} (J \rho v^{x}) + \partial_{y} (J \rho v^{y}) + \partial_{\tilde{z}} (J\rho v^{\tilde{z}}) = 0 \quad .
\end{equation}

In an ideal, barotropic fluid vorticity is a conserved tensor density which is related to the relabelling symmetries of ideal fluid Lagrangian \citep{Padhye96}. The local model preserves this relabelling symmetry so we should expect the model to have a form of vorticity conservation. In the global coordinates the fluid vorticity obeys

\begin{align}
\begin{split}
 \mathcal{D} \omega^i &=  D \omega^i - \omega^{j} \nabla_{j} u^i + \omega^i \nabla_j u^{j} , \\
&=  (\partial_t  + u^i \partial_i) \omega^i - \omega^{j} \partial_{j} u^i + \omega^i \nabla_j u^{j}  , \\
&= 0 ,
\end{split} \label{global vorticity}
\end{align}
where, in this paragraph only, $D = \partial_t  + u^i \nabla_i$ is the Lagrangian time derivative with respect to the total (global+local) flow. The latter expression being a consequence of the symmetry of the Christoffel symbols. The vorticity depends only on the relative velocity as the background flow is irrotational,

\begin{equation}
\omega^{i} = \varepsilon^{i j k} \partial_{j} v_k \quad ,
\end{equation}
where $\varepsilon^{i j k}$ is the volume element of the local model. Substituting Equation \ref{background local split}, into Equation \ref{global vorticity}, for the fluid velocity we obtain, in local (unstretched) coordinates,

\begin{align}
\begin{split}
  D \omega^i &+ \omega^i [\Delta+ \partial_{x} v^{x} + \partial_{y} v^{y} + \partial_{z} v^{z}] - \omega^{z} U_{R 0} \delta^{i}_{z} \\
&- (\omega^{x} \partial_{x} + \omega^{y} \partial_{y} + \omega^{z} \partial_{z} ) v^{i} = 0 \quad .
\end{split}
\end{align}
Upon introducing the stretched vertical coordinate, $\tilde{z}$, this simplifies to

\begin{equation}
  D \omega^i + \omega^i [\Delta+ \partial_{x} v^{x} + \partial_{y} v^{y} + \partial_{\tilde{z}} v^{\tilde{z}}] - (\omega^{x} \partial_{x} + \omega^{y} \partial_{y} + \omega^{\tilde{z}} \partial_{\tilde{z}} ) v^{i} = 0\, .
\end{equation}
It is informative to write the above in terms of the tensor density advection operator with respect to the relative flow, $\mathcal{D}_{\rm rel}$ ,

\begin{equation}
 \mathcal{D}_{\rm rel} \omega^i = - \omega^i \Delta \quad ,
 \label{vorticity equation}
\end{equation}
where, when acting on the vorticity, the advection operator with respect to the background flow is

\begin{equation}
 \mathcal{D}_{\rm rel} \omega^i = D \omega^i + \omega^i \partial_j v^{j} - \omega^{j} \partial_{j} v^i ,
\end{equation}
with $D = \partial_{\tau} + v^i \partial_i$ the usual Lagrangian time derivative in the local model. Thus we see that the change in box volume, through $\Delta$, results in a source/sink of vorticity. In the global picture the vorticity is advected by the total flow, in the local picture the vorticity is advected by the relative flow with the background flow appearing as a source/sink. This is analogous to the situation in cosmology where the peculiar velocities redshift to zero in an expanding FRW metric resulting in $\omega_i \propto a^{-1}$ \citep{Mo10}. The closely related quantity, $J \omega^{i}$, is conserved by the local model, with no source/sink from the box volume change. The effect of the changing box volume on the fluid vorticity can be seen in the simulations of \citet{Robertson12} (where it is referred to as adiabatic heating), with a rapidly contracting box leading to a strengthening of large scale eddies and a corresponding increase in the turbulent velocities. 

The kinetic helicity, $H_{k}$, is a conserved quantity associated with the vorticity which is conserved in barotropic ideal fluids. Despite the collapse/expansion acting as a source/sink of vorticity, one can show that the kinetic helicity of the relative flow is conserved. The kinetic helicity of the relative flow is

\begin{align}
\begin{split}
 H_k &= \iiint v_i \varepsilon^{i j k} \partial_{j} v_{k} d V , \\
        &= \iiint v_i \varepsilon^{i j k} \partial_{j} v_{k} J d x \, d y \, d \tilde{z} . \\
\end{split}
\end{align}
Taking the time derivative of $H_k$,

\begin{align}
\begin{split}
 \dot{H}_k &= \iiint \left( \dot{v}_i \varepsilon^{i j k} \partial_{j} v_{k} + v_i \varepsilon^{i j k} \partial_{j} \dot{v}_{k}\right) J d x \, d y \, d \tilde{z} , \\
&= -\iiint \Biggl[ \left( v^{a} \partial_{a} v_{i}  + \partial_i h \right) \varepsilon^{i j k} \partial_{j} v_k \\
&+ v_i \varepsilon^{i j k} \partial_{j} \left( v^{a} \partial_{a} v_{k} + \partial_k h  \right) \Biggr] J d x \, d y \, d \tilde{z} , \\
&= -\iiint  \partial_{j} \left( \varepsilon^{i j k} v_k v^{a} \partial_{a} v_{i} + \varepsilon^{i j k} v_k \partial_i h \right) J d x \, d y \, d \tilde{z} , \\
&= 0 . 
\end{split}
\end{align}
where we have made use of the antisymmetry of $\varepsilon^{i j k}$, $\partial_t (\varepsilon^{i j k} J) = 0$, along with the periodic boundary conditions. Conservation of kinetic helicity in the local model is not surprising as the stretching due to the background flow is not able to change the flow topology in the box.

Unlike hydrodynamics in both Cartesian and spherical geometries, the local model does not posses time translation symmetry. One can still obtain an energy equation for the relative motion. The conservative form of the energy equation for the relative motion in the, stretched, local coordinates is

\begin{align}
\begin{split}
 \partial_{\tau} \left( J \rho \mathcal{E}_{\rm rell}  \right) + \partial_{i} \left[ J v^i \left( \rho  \mathcal{E}_{\rm rell}  + p \right) \right] = \\
- J \rho \left[ U_{R} L_z^2 v^{\tilde{z}} v^{\tilde{z}}  + \frac{U_0}{\mathcal{R}} \mathcal{R}^2  \left( v^x v^x + v^y v^y \right) \right] - J p \Delta ,
\end{split}
\end{align}
where the energy of the relative motion is given by

\begin{equation}
\mathcal{E}_{\rm rell} = \frac{1}{2} L_{z}^2 v^{\tilde{z}} v^{\tilde{z}} + \frac{1}{2} \mathcal{R}^2 (\tau) [v^{x} v^{x} + v^{y} v^{y}] + \varepsilon 
\end{equation}
Thus we see that the background flow is a source/sink of energy for the local model, through the $p d V$ work done by the flow, and through the gradients in the background flow which are accessed through the Reynolds stresses. This is similar to the energy equations obtained in distorted shearing box models \citep[e.g.][]{Ogilvie14}

\section{Nonlinear-Solutions} \label{nonlinear solutions}

\subsection{Horizontal shear flows} \label{horizontal shear flows}

We can consider solutions to the local model with a spatially homogeneous density $\rho = \rho (\tau)$ and a purely horizontal velocity field ($v^{z} = 0$) of the form

\begin{equation}
 v^{i} = A(t) v_0 (\tilde{z}, \lambda) \hat{v}^{i} (\tilde{z}) ,
\label{horizontal shear solution}
\end{equation}
where $\lambda = \hat{v}^{y}(\tilde{z}) x - \hat{v}^{x} (\tilde{z}) y$. This flow consists of a horizontal shear flow that is uniform along the flow direction, $\hat{\mathbf{v}}$, but can vary perpendicular to $\hat{\mathbf{v}}$. This result in shear across these perpendicular directions. We set $A=1$ at some initial time $\tau=\tau_0$, so that $v_0 (\tilde{z}, \lambda ) \hat{v}^{i} (\tilde{z})$ represents the initial velocity field in the fluid. 

Note that we have

\begin{align}
\begin{split}
 \partial_{x} v^{x} + \partial_{y} v^{y} &= A \left( \hat{v}^{x} \partial_{x} v_0 + \hat{v}^{y} \partial_{y} v_0 \right) \\
&= A \left( \hat{v}^{x} \hat{v}^{y} \partial_{\lambda} v_0 - \hat{v}^{y} \hat{v}^{x}  \partial_{\lambda}  v_0 \right) \\
&= 0 ,
\end{split} 
\end{align}
meaning this velocity field doesn't cause a change in density, or pressure. Also, we have

\begin{align}
\begin{split}
 D v_0 &= A v_0 \left( \hat{v}^{x} \partial_{x} v_0 + \hat{v}^{y} \partial_{y} v_0 \right) \\
&= A v_0 \left( \hat{v}^{x} \hat{v}^{y} \partial_{\lambda} v_0 - \hat{v}^{y} \hat{v}^{x}  \partial_{\lambda} v_0 \right) \\
& = 0 .
\end{split} 
\end{align}
Given the above, Equation \ref{horizontal shear solution} is a solution to the local model provided that the amplitude $A$ evolves according to

\begin{equation}
\dot{A} + \frac{2 U_0}{\mathcal{R}} A .
\end{equation}
Making use of $d \mathcal{R}/d \tau = U_0$ we find that,

\begin{equation}
A = \left(\mathcal{R}/R_0\right)^{-2} ,
\end{equation}
yielding the following for the velocity field at time $t$,

\begin{equation}
 v^{i} = v_0 (\tilde{z}, \hat{v}^{y}(\tilde{z}) x - \hat{v}^{x} (\tilde{z}) y ) \hat{v}^{i} (\tilde{z}) \left(\mathcal{R}/R_0\right)^{-2}  .
\end{equation}
This is the local manifestation of conservation of angular momentum in the zonal flows of the global model.

\begin{figure}
\includegraphics[trim=100 420 0 150, clip, width=\linewidth]{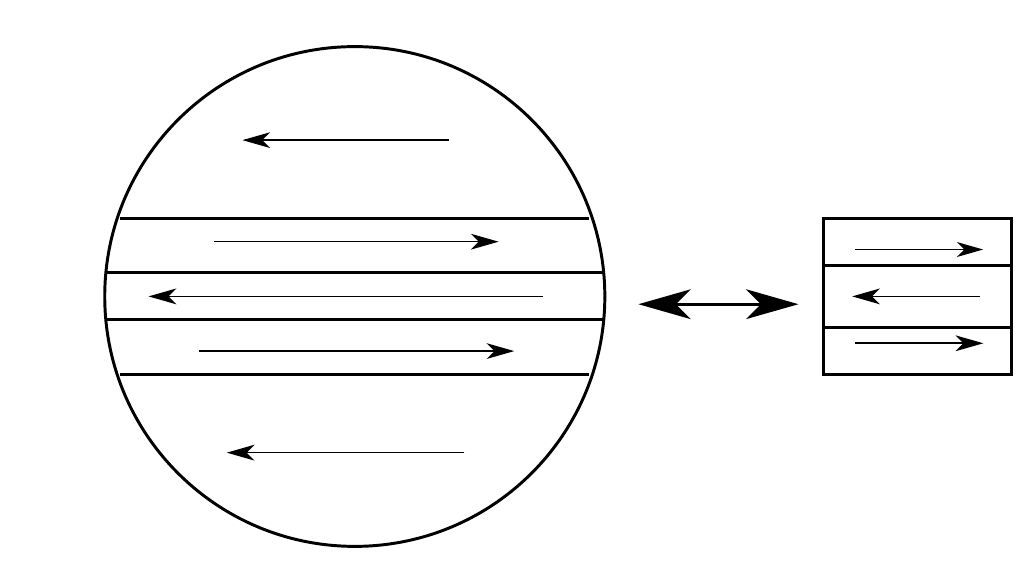}
\caption{Horizontal shear flows in the local model are equivalent to (uniform density) zonal flows on the collapsing(/expanding) reference sphere in the global model.}
\label{zonal flow cartoon}
\end{figure}

These represent stratified zonal flows, of constant density, in the global geometry (as illustrated in Figure \ref{zonal flow cartoon}) which grow/attenuate due to conservation of angular momentum with a change in the size of the reference sphere. The orientation of these zonal flows is free to vary with height due to the absence of shear stress.

In a collapse these zonal flows grow in strength with time. It is unlikely that the shear flow can steepen indefinitely without going unstable, likely by going Kelvin-Helmholtz unstable at the locations of maximum shear. This may result in zonal flows of approximately constant velocity separated by horizontal vortices/rolls. As the collapse proceeds still further these zonal flows will continue to strengthen. Whether the nonlinear outcome of this collapse results in a disruption of the ordered vortex layer, zonal flow geometry into fully developed turbulence throughout the box, or in a reorientation of the zonal flows until they self-organise into a stable shear flow will require numerical simulation to determine. 

Of course if the collapse is allowed to proceed indefinitely the zonal flows will become so strong that they break the asymptotic scheme used to derive the local model. In the global model these zonal flows will ultimately grow strong enough to provided rotational support to the gas, slowing or halting the collapse.

\subsection{Elevator Flows}

A second class of nonlinear flows which can be supported by the local model are vertically homogeneous `elevator flows' of constant density. These can only occur for periodic vertical boundaries, an appear to be analogous to the elevator flows seen in some unstratified disc simulations \citep[e.g.][]{Dewberry19,Dewberry20}. Here the density, and pressure, is a function of time only $\rho = \rho(\tau)$. The horizontal velocities are zero, $v^{x} = v^{y} = 0$. The vertical velocity is given by,

\begin{equation}
 v^{\tilde{z}} = B (\tau) v_0^{\tilde{z}} (x, y) .
 \label{elevator flow}
\end{equation}
Where $v_0^{\tilde{z}}$ is the vertical velocity at $\tau = \tau_0$. This works because $\partial_{\tilde{z}} v^{\tilde{z}} = 0$, meaning this velocity field results in no change in density, and $D v_0^{\tilde{z}} = v^{\tilde{z}} \partial_{\tilde{z}}^{\prime} v_0^{\tilde{z}} = 0$. Substituting Equation \ref{elevator flow} into the vertical momentum equation we obtain the following equation for the evolution of $B$,

\begin{equation}
\dot{B} + 2 B U_{R 0} = 0 .
\end{equation}
Making use of $d L_z/d \tau = U_{R 0} L_z$ we find that,

\begin{equation}
 B = (L_z/L_{z 0})^{-2} ,
\end{equation}
yielding the following for the velocity field at time $t$,

\begin{equation}
 v^{\tilde{z}} =  v_0^{\tilde{z}} (x, y) (L_z/L_{z 0})^{-2}  .
\end{equation}
This grows provided that $U_{R 0} < 0$.

These elevator flows can only occur with vertically periodic boundaries, and do not occur for closed/reflective boundary conditions. Unlike the horizontal flows considered in Section \ref{horizontal shear flows}, which are the local realisation of zonal flows present in the global geometry, these vertical elevator flows may be an artefact of periodic boundaries. Similar elevator flows are found in simulations of accretion discs with vertically periodic boundaries, despite not being present in the global problem \citep{Dewberry19,Dewberry20}.

\subsection{Diagonal flows} \label{diagonal flows}

\begin{figure}
\includegraphics[trim=100 420 0 150, clip, width=\linewidth]{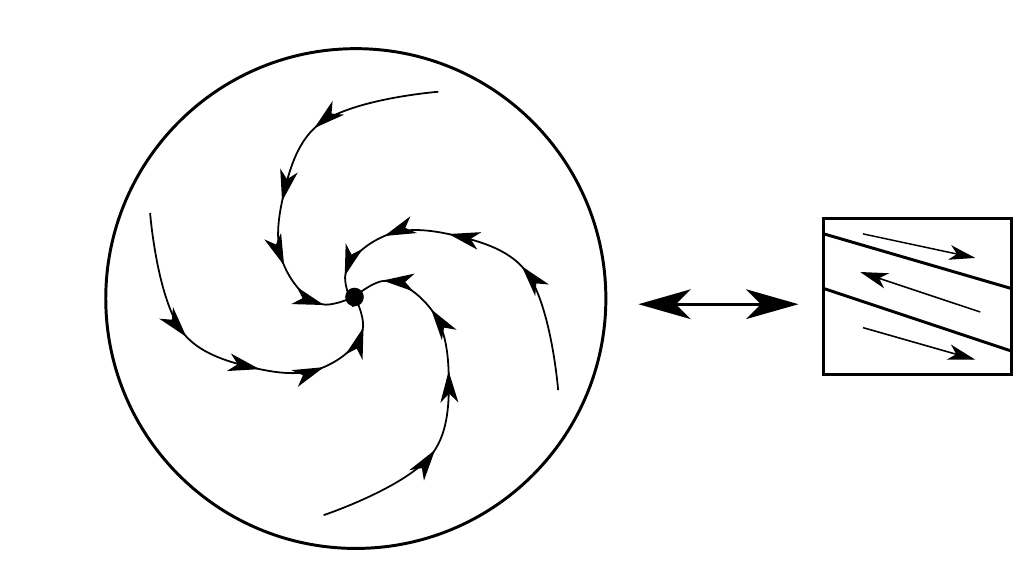}
\caption{More generally these uniform density shear flows in the local model are local representations of a (weakly) differentially rotating flow with the streamlines in the global model spiralling towards/away from the centre of the sphere.}
\label{spiral flow cartoon}
\end{figure}

There are a set of diagonal flows of the form

\begin{align}
 v^x &= \left(\mathcal{R}/R_0\right)^{-2} v_0^{x} (y) , \\
 v^{\tilde{z}} &= \left(L_{z}/L_{z 0}\right)^{-2} v_0^{\tilde{z}} (y) ,
\end{align}
where, without loss of generality we can align the diagonal flow with the $x$-axis. Globally these flows represent a uniform density spiral in the global flow (as illustrated in Figure \ref{spiral flow cartoon}) where the pitch angle of the spiral varies with `latitude' on the reference sphere. As the vertical and horizontal velocities evolve separately the pitch angle of the spiral changes with time.

In general the horizontal shear, elevator and diagonal flows are local representations of weak departures from spherical symmetry in the global model. This is similar to the local representation of disc warps and eccentricity in (circular) shearing box (e.g. \citet{Balbus99,Latter06,Ogilvie22}).

\subsection{Other nonlinear solutions} 

It is possible to find additional nonlinear solutions; notably it appears to be possible  to generalise homentropic/Kidda-like vortex solution for certain background flows. Additionally adopting shear periodic boundary allows for a more general class of shear flow than those considered here. We leave exploration of both of these classes of solution to future work as they are of interest in their own right.

\section{Linear perturbations} \label{linear perturbations}

We now explore the behaviour of linear waves in the local model. For our background state we take $\rho = \rho(\tau)$ and $v^i = 0$, where

\begin{equation}
 D \rho = - \rho \Delta ,
\end{equation}
with $\Delta$ the velocity divergence of the background flow. The linearised equations of motion and continuity, on top of this background state, are

\begin{align}
 D \delta v^x + \frac{2 U_0}{\mathcal{R}} \delta v^x  &= - \mathcal{R}^{-2} \partial_x \delta h \\
 D \delta v^y + \frac{2 U_0}{\mathcal{R}} \delta v^y  &= - \mathcal{R}^{-2} \partial_y \delta h \\
 D \delta v^{\tilde{z}} + 2 U_{R 0} \delta v^{\tilde{z}}  &= - L_z^{-2} \partial_{\tilde{z}} \delta h \\
 D \delta \rho &= - \rho \partial_i \delta v^{i} - \delta \rho \Delta .
\end{align}
The enthalpy perturbation, $\delta h$, can be written in terms of the relative density perturbation $\delta \psi = \frac{\delta \rho}{\rho}$ through,

\begin{equation}
 \delta h = c_s^2 \delta \psi .
\end{equation}
The relative density perturbation evolves according to

\begin{equation}
 D \delta \psi = - \partial_i \delta v^{i}  .
\end{equation}
Introducing a Lagrangian displacement \citep[following][]{Lynden-Bell67,Papaloizou05a} of the form $\boldsymbol{\xi} = \hat{\boldsymbol{\xi}} \exp (i \mathbf{k} \cdot \mathbf{x})$ such that $\delta \mathbf{v} = D \boldsymbol{\xi}$ and the relative density perturbation which is related to the Lagrangian displacement by

\begin{equation}
\delta \hat{\psi} = -i \mathbf{k} \cdot \hat{\boldsymbol{\xi}} .
\end{equation}
Substituting these into the linearised momentum equation we obtain

\begin{align}
 D^2 \hat{\xi}^{x} + \frac{2 U_0}{\mathcal{R}} D \hat{\xi}^{x} &= -c_{s H}^2 k_x k_i \hat{\xi}^{i} , \label{linear x eq} \\
 D^2 \hat{\xi}^{y} + \frac{2 U_0}{\mathcal{R}} D \hat{\xi}^{y} &= -c_{s H}^2 k_y k_i \hat{\xi}^{i} , \\
 D^2 \hat{\xi}^{\tilde{z}} +  2 U_{R 0} D \hat{\xi}^{\tilde{z}} &= -c_{s V}^2 k_{\tilde{z}} k_i \hat{\xi}^{i} . \label{linear z eq} 
\end{align}
The can be derived from the following Lagrangian density (which can alternatively be obtained from Equation \ref{nonlinear Lagrangian})

\begin{equation}
 \mathcal{L} = \frac{1}{2} \mathcal{R}^2 (\delta \hat{v}^x \delta \hat{v}^x + \delta \hat{v}^y \delta \hat{v}^y) + \frac{1}{2} L_z^2 \delta \hat{v}^{\tilde{z}} \delta \hat{v}^{\tilde{z}} - \frac{1}{2} c_s^2 (k_i \hat{\xi}^i)^2 ,
\end{equation}
where, without loss of generality we can rotate the box such that $k_y = 0$. Variation with respect to $\hat{\boldsymbol{\xi}}$ leads to Equations \ref{linear x eq}-\ref{linear z eq}. Associated with this Lagrangian we have the Hamiltonian density

\begin{equation}
\mathcal{H} = \frac{1}{2} \mathcal{R}^{-2} (\hat{\pi}_x^2 + \hat{\pi}_y^2 + b^{-2} \hat{\pi}_{\tilde{z}}^2 ) + \frac{1}{2} c_s^2 (k_i \hat{\xi}^i)^2 ,
\label{linear Hamiltonian}
\end{equation}
where $\hat{\pi}_x = \mathcal{R}^2 \delta \hat{v}^x$, $\hat{\pi}_y = \mathcal{R}^2 \delta \hat{v}^y$, $\hat{\pi}_{\tilde{z}} = L_z^2 \delta \hat{v}^{\tilde{z}}$ and we have made use of the aspect ratio $b = L_{z}/\mathcal{R}$. Performing a canonical transform using the following generating function

\begin{equation}
 G = P_{\alpha} k_i \hat{\xi}^{i} + P_{\beta} (\aleph \hat{\xi}^{x} + \beth \hat{\xi}^{\tilde{z}} ) + P_{\gamma} \hat{\xi}^{y} ,
\end{equation}
where $\aleph$ and $\beth$ \footnote{The Hebrew letters aleph and beth} are to be determined. This results in the following relationship between old and new coordinates and momenta,

\begin{align}
 \hat{\pi}_x &= P_{\alpha} k_{x} + P_{\beta} \aleph \\
 \hat{\pi}_{y} &= P_{\gamma} , \\
 \hat{\pi}_{\tilde{z}} &= P_{\alpha} k_{\tilde{z}} + P_{\beta} \beth \\
 \alpha &= k_i \hat{\xi}^{i} , \\
 \beta &= \aleph\hat{\xi}^{x} + \beth \hat{\xi}^{\tilde{z}} , \\
 \gamma &= \hat{\xi}^{y} ,
\end{align}
Consider the contribution to the Hamiltonian density from $\hat{\pi}_x^2 + b^{-2} \hat{\pi}_{\tilde{z}}^2$. Writing this in terms of the new momenta we obtain

\begin{align}
\begin{split}
 \hat{\pi}_x^2 + b^{-2} \hat{\pi}_{\tilde{z}}^2 &= (P_{\alpha} k_{x} + P_{\beta} \aleph)^2 + b^{-2} (P_{\alpha} k_{\tilde{z}} + P_{\beta} \beth)^2 \\
&= (k_{x}^2 + b^{-2} k_{\tilde{z}}^2) P_{\alpha}^2 + (\aleph^2 + b^{-2} \beth^2) P_{\beta}^2 \\
&+ 2 (\aleph k_x + k_{\tilde{z}} b^{-2} \beth) P_{\alpha} P_{\beta} .
\end{split}
\end{align}
For boxes of constant aspect ratio we can diagonalise this by setting $\aleph k_x + k_{\tilde{z}} b^{-2} \beth = 0$, meaning the $\alpha$, $\beta$ and $\gamma$ linear perturbations are decoupled. In general choosing 

\begin{align}
 \aleph &= \frac{k_{\tilde{z}}}{\sqrt{k_x^2 b_0^2 + k^2_{\tilde{z}}}} ,\\
 \beth  &= -\frac{k_x b_0^2}{\sqrt{k_x^2 b_0^2 + k^2_{\tilde{z}}}} ,
\end{align}
where we have introduced $b_0 = b(0)$, means that the dynamics of the $\alpha$, $\beta$ and $\gamma$ waves are initially independent, but changes to the box aspect ratio introduces a `mixing' of the $\alpha$ and $\beta$ waves. With this choice this leads to the following, transformed, Hamiltonian density:

\begin{align}
\begin{split}
 \mathcal{H} &= \frac{1}{2} \mathcal{R}^{-2} \Biggl\{  (k_{x}^2 + b^{-2} k_{\tilde{z}}^2) P_{\alpha}^2 + \frac{k_x^2 b_0^2 (b/b_0)^{-2} + k_{\tilde{z}}^2}{k_x^2 b_0^2 + k^2_{\tilde{z}}} P_{\beta}^2 \\
&+ 2 \frac{k_{\tilde{z}} k_x}{\sqrt{k_x^2 b_0^2 + k^2_{\tilde{z}}}} [1- (b/b_0)^{-2} ] P_{\alpha} P_{\beta}  +  P_{\gamma}^2  \Biggr \} + \frac{1}{2} c_s^2 \alpha^2 .
\end{split}
\end{align}

This Hamiltonian is independent of both $\beta$ and $\gamma$, meaning $P_{\beta}$ and $P_{\gamma}$ are both constants of motion and $\beta$ and $\gamma$ satisfy

\begin{align}
 \dot{\beta} &= \mathcal{R}^{-2}  \frac{k_x^2 b_0^2 (b/b_0)^{-2} + k_{\tilde{z}}^2}{k_x^2 b_0^2 + k^2_{\tilde{z}}} P_{\beta} + \mathcal{R}^{-2}  k_{\tilde{z}} k_x \frac{1- (b/b_0)^{-2} }{\sqrt{k_x^2 b_0^2 + k^2_{\tilde{z}}}} P_{\alpha}  \label{beta evo} , \\
 \dot{\gamma} &= \mathcal{R}^{-2} P_{\gamma} .
\end{align}
The latter just being a linear shear flow in the $y-$direction. $P_{\beta}$ and $P_{\gamma}$ can be related to the Fourier components of the vorticity perturbation, $\delta \omega^i$, through

\begin{align}
 \delta \omega^{x} &= - \frac{i k_{\tilde{z}}}{J} P_{\gamma} , \\
 \delta \omega^{x} &= - \frac{i}{J} \sqrt{k_x^2 b_0^2 + k_{\tilde{z}}^2} P_{\beta} , \\
 \delta \omega^{\tilde{z}} &= \frac{i k_{x}}{J} P_{\gamma} .
\end{align}
The linearised form of Equation \ref{vorticity equation} is $\partial_t (J \delta \omega^{i}) = 0$, so we see that $P_{\beta}$ and $P_{\gamma}$ being integrals of motion arises as a result of vorticity conservation. We can thus identify the $(\alpha,P_{\alpha})$ with the sound wave, while $(\beta,P_{\beta})$ and $(\gamma,P_{\gamma})$ are vortical waves. Changes in the box aspect ratio leads to coupling between the sound wave and the $(\beta,P_{\beta})$ vortical waves.

Treating $P_{\beta}$ and $P_{\gamma}$ as parameters, the Hamiltonian for the dynamics of $(\alpha, P_{\alpha})$ reduces to

\begin{equation}
 \mathcal{H} = \frac{1}{2} \mathcal{R}^{-2}  (k_{x}^2 + b^{-2} k_{\tilde{z}}^2) P_{\alpha}^2  + \mathcal{R}^{-2} k_{\tilde{z}} k_x \frac{1- (b/b_0)^{-2}}{\sqrt{k_x^2 b_0^2 + k^2_{\tilde{z}}}} P_{\alpha} P_{\beta} + \frac{1}{2} c_s^2 \alpha^2 . \label{sound wae hamiltonian}
\end{equation}
One can show that the dynamics of this system correspond to a forced harmonic oscillator with a variable frequency by switching the position and momenta,

\begin{equation}
 \mathcal{H} = \frac{1}{2} \Pi^2 + \frac{1}{2} \omega^2 X^2 - g P_{\beta} X ,
\end{equation}
where $\Pi = c_s \alpha$, $X = -c_s^{-1} P_{\alpha}$ and we have introduced the sound wave frequency, $\omega$, and coupling coefficient, g,  

\begin{align}
\omega &= c_s |\mathbf{k}| = c_s (\mathcal{R}^{-2} k_x^2 + L_{z}^{-2} k^2_{\tilde{z}} )^{1/2} , \label{sound wave freq} \\
g &=  c_s \mathcal{R}^{-2} k_x k_{\tilde{z}} \frac{1- (b/b_0)^{-2}}{\sqrt{k_x^2 b_0^2 + k^2_{\tilde{z}}}} . \label{coupling coefficient}
\end{align}

There are two regimes depending on whether the waveperiod is short relative to $t_{\rm bg}$. When $\omega t_{\rm bg} \gg 1$ we have the WKB/modulated wave regime where the linear wave consists of a simple wave propagating on a slowly varying background. Taking $\omega$ as a large parameter, with $g = O(\omega^2)$, and provided $P_{\beta} = O(\omega^{-2})$ then the WKB solution for $X$ is

\begin{equation}
 X = X_{\pm} \omega^{-1/2} \exp \left( \pm i \int \omega d \tau\right) .
\end{equation}
The WKB solution for $\alpha$, $P_{\alpha}$ are then

\begin{align}
  \alpha &\approx \pm i c_s^{-1} X_{\pm} \omega^{1/2} \exp \left( \pm i \int \omega d \tau\right) , \label{alpha WKB} \\
 P_{\alpha} &\approx - c_s X_{\pm} \omega^{-1/2} \exp \left( \pm i \int \omega d \tau\right) ,
\end{align}
while $P_{\beta}$ is a (small) constant. The WKB solutions, specifically Equation \ref{alpha WKB} which is related to the density perturbation, show that if $\omega \rightarrow \infty$ at late times, for some choice of $k_x/k_{\tilde{z}}$ then there exists perturbations to the basic state (sound waves) which can grow to large amplitude. For horizontally propagating waves this corresponds to $\mathcal{R} \rightarrow 0$ (at late times), while for vertically propagating waves this is $L_{z} \rightarrow 0$. This condition is not sufficient, however, to determine if the nonlinear saturation of such growing perturbations allows them to attain sufficient amplitude to affect the background flow.

In the opposite regime $\omega t_{\rm bg} \lesssim 1$ we have the freeze out regime where the wave phase cannot undergo a full cycle on the timescale of the background flow and the wave becomes a roughly static pattern which is deformed by the background flow. These two regimes (modulated wave v.s. freeze out), seen in linear waves, has similarly been found for turbulence in contracting boxes \citep{Robertson12}, where a rapidly contracting box leads to an increase in the turbulent velocities on a timescale too short for energy to be passed down the turbulent cascade.

\begin{figure}
\includegraphics[trim=0 0 40 40, clip, width=\linewidth]{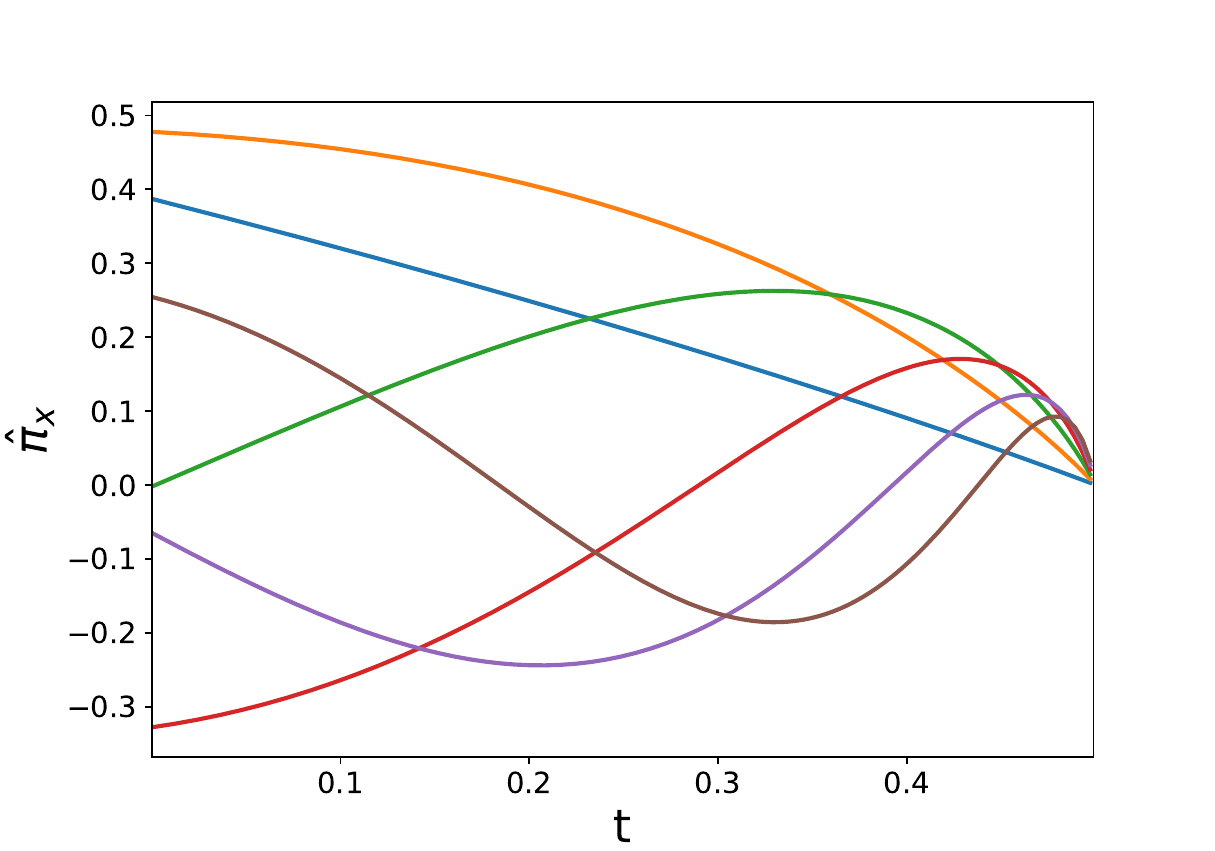}
\caption{Amplitude (as measured by $\hat{\pi}_x$) of a horizontally propagating sound waves for a freely falling box with $\mathcal{R} = R_0 (1 - t/t_c)^{2/3}$, $R_0 = 1$, $t_c = 0.5$ and wavenumbers $k_x = n c_{s 0}^{-1}$ where $n$ is an integer. The solutions correspond to Bessel functions (See Appendix \ref{beyond WKB}). In the freeze out regime ($\omega t_{\rm bg} \lesssim 1$) the solution has no nodes on the timescale of the collapse.}
\label{free fall sound}
\end{figure}

\begin{figure}
\includegraphics[trim=0 00 40 40, clip, width=\linewidth]{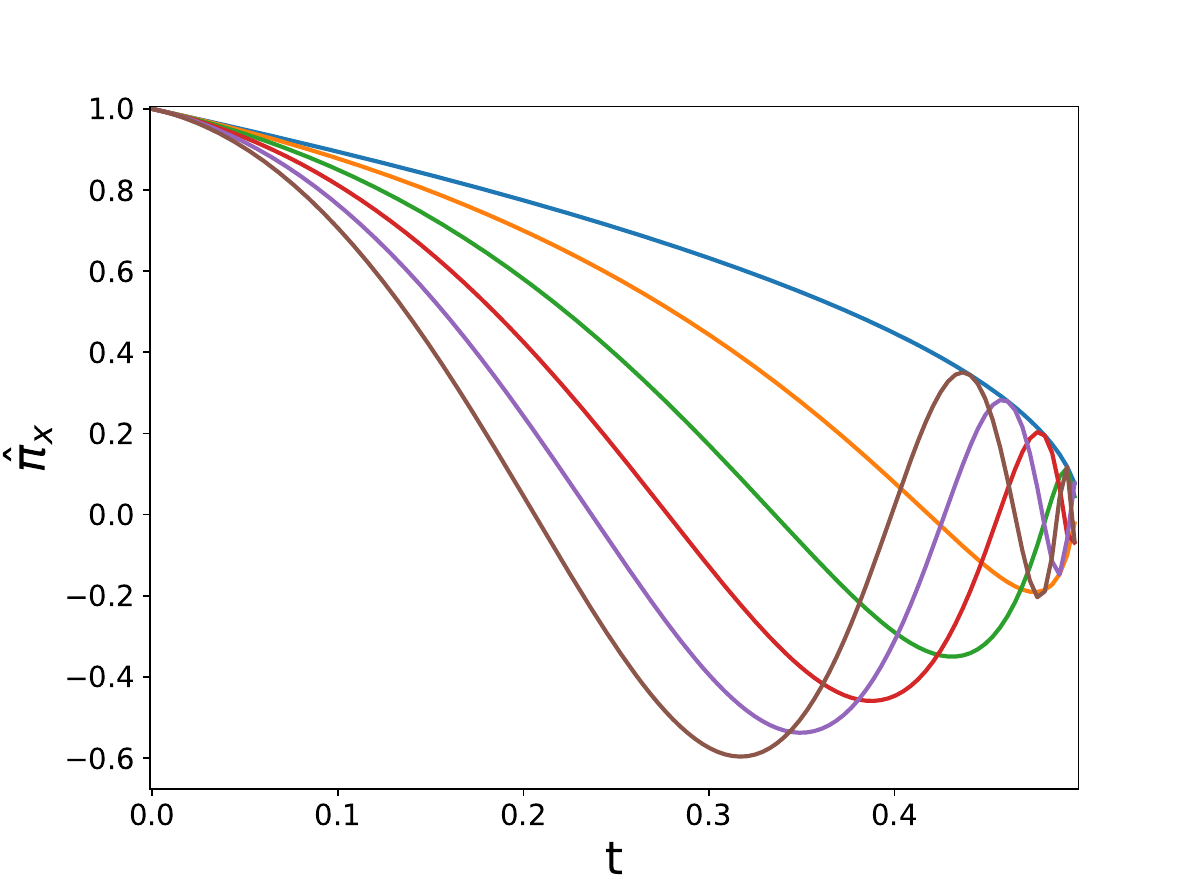}
\caption{Amplitude of a horizontally propagating sound waves for a uniform collapse with $\mathcal{R} = R_0  + U_0 t$, $R_0 = 1$, $U_0 = 2$ and wavenumbers $k_x = n c_{s 0}^{-1}$ where $n$ is an integer.  The solution corresponding to (complex) powerlaws. In the freeze out regime the powerlaw exponent is strictly real and there are no oscillations of the wave.}
\label{uniform sound}
\end{figure}

A more detailed analysis of the linear waves in the local model is presented in Appendix \ref{linear wave analysis}. Appendix \ref{beyond WKB} derives exact solutions for sound waves which are decoupled from the vortical wave (i.e. $g=0$), for various background flows. Figures \ref{free fall sound} and \ref{uniform sound} show example solutions for horizontally propagating sound waves for free fall ($\mathcal{R} = R_0 (1 - t/t_c)^{2/3}$) and uniform collapse ($\mathcal{R} = R_0  + U_0 t$) profiles respectively. In both cases the solution possesses no nodes on the timescale of the collapse in the freeze out regime. The uniform collapse profile provides a useful illustration of the transition to the freeze out regime as $P_{\alpha}$ takes the form $P_{\alpha} \propto \mathcal{R}^{1/2 \pm \lambda}$ where $\lambda = \frac{1}{2} \sqrt{1 - 4 (c_S k_x/U_0)^2}$. For $k_x > \frac{1}{2} \frac{U_0}{c_s}$, $\lambda$ is imaginary and we have a left and right propagating sound waves. While for $k_{x} < \frac{1}{2} \frac{U_0}{c_s}$, $\lambda$ is real and the left and right sound waves are frozen out, becoming two non-propagating waves with differing growth rates.

\begin{figure}
\includegraphics[trim=150 520 180 100, clip, width=\linewidth]{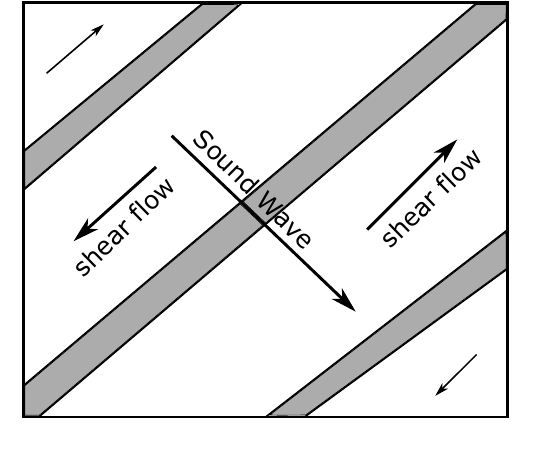}
\caption{Cartoon illustrating the relationship between the diagonal sound waves and shear flows. The shear flow is generated parallel to the direction of the sound wave propagation. Likewise a diagonal shear flow will generate bands in the density running parallel to the shear flow as a result of exciting a standing sound wave}
\label{diagonal flow}
\end{figure}

Appendix \ref{diagonally propergating waves} explores the diagonal linear waves (i.e. with $k_x$, $k_{\tilde{z}} \ne 0$) in more detail. Notably for diagonal waves the sound wave ($P_{\alpha}$, $\alpha$) and vortical waves ($P_{\beta}$, $\beta$) are coupled as a result of the changing aspect ratio of the box. This means a diagonally propagating sound wave will generate a shear flow perpendicular to the direction of wave propagation (illustrated in figure \ref{diagonal flow}). Likewise an initially uniform density diagonal shear flow will generate a standing sound wave perpendicular to the fluid motion resulting in bands of varying density running parallel to the shear flow which arise. One complicating factor is in constant aspect ratio models, with $b\ne1$, diagonally propagating sound waves, will produce shearing motions in the local model which are purely a result of the choice of coordinate system and do not represent physical shearing motions in the fluid. For numerical implementations, these sound wave generated shearing motions (either physical or as a result of the choice of coordinates) may present a challenge to some Riemann solvers.

In the global picture the coupling occurs as a result of the advection of the background flow by the velocity perturbations in the diagonal waves.  In many collapse profiles diagonal flows tend to become more horizontal with time. Thus one expects a weakly rotating collapse (which can be represented with a diagonal shear flow in the local model) to spontaneously generate tightly wound density waves at late times.

\section{Discussion} \label{discussion}

\subsection{Implementing the local model in hydrocodes}

{The utility of a local model will be dependant on a practical numerical implementation. This has hindered the adoption of some local models in the past, notably the distorted shearing boxes with their time dependant geometries, and one is often left with the choice of attempting to coerce an existing code into running with an unusual geometry, or writing an entire specialised hydrocodes just for this specific application. These problems affects even relatively commonly used local models such as the regular shearing box, in particular with how to implement shear periodic boundaries. Examples of specific (magneto-)hydrodynamic codes being developed to model specific local model such as \citet{Paardekooper19} for the warped shearing box, \citet{Wienkers18} for the eccentric shearing box and \citet{Shi20,Huang22} for the expanding box. An alternative approach was pursued by \citet{Latter06} and \citet{Ogilvie22} (for eccentric and warped disc respectively) who instead made use of a regular shearing box but used a local representation of the distorted flow in setting up their model - which corresponds to the long wavelength part of the fluid flow on the box lengthscale. This latter approach only works if the distorted and undistorted flows are similar enough that the distorted flow fits into the box (i.e. this works for linear warps/eccentricity). Appendix \ref{alternative formalisms} presents several alternative formulations of the local model which may be useful for incorporating the model into existing hydrocodes.

\textbf{\subsection{On self gravity in the local model}}

In this paper we have chosen not to consider self-gravity in our local model. It is worth noting that we have not made use of an explicit closure for the background gravitational potential $\Phi$, which is only constrained to be spherically symmetric and varying on a lengthscale much longer than $L_H$ and $L_V$. Thus our local model is compatible with a self-gravitating background flow provided that the self gravity of the density perturbations within the box can be ignored.

Looking ahead it seems likely one can include the self gravity of the local flow by solving the Poisson's equation for the density difference $\rho - \rho_0$ subject to periodic boundaries. $\rho - \rho_0$ has zero mean, thus Poisson's equation can be inverted in Fourier space and the solutions likely correspond to the high degree spherical harmonics of the gravitational potential of the global problem. We leave demonstrating this, and exploring the effects of self gravity, to future work.

\subsection{Extensions}

While interesting in it's own right, there are a number of useful generalisations to the local model that could be carried out. The obvious extension, given the close relationship between our model and the EBM/AEB, is to extend the model to include magnetic fields; likely following a similar approach to the MHD eccentric shearing box \citep{Ogilvie14} rather than the approach of \citet{Tenerani17} for consistency with how we treat pressure. Regardless of the approach it is worth checking whether including gradients in the background magnetic field (as done by \citet{Tenerani17}) will lead to spurious instabilities similar to gradients in the background pressure. We have chosen not to consider magnetic fields here in order to focus on the properties of the hydrodynamic model, particularly during collapse which has not been looked at before. Additionally while adding magnetic fields to the local model in an expanding flow is straight forward, for a contracting flow one encounters additional complications. Magnetic fields in collapses are known to break spherical symmetry \citep{Galli93,Hennebelle01,Galli06} meaning one would need to make significant modifications to both the background flow and the resulting geometry of the local model to include them. 

The second important generalisation, particularly for realistic collapse profiles, is to include rotation in the background flow \citep[such as][]{Hennebelle01,Hennebelle03,Galli06}. As discussed in Section \ref{nonlinear solutions}, slowly rotating collapses can already be modelled in our existing local model and take the form of growing, horizontal, shear flows in the local model. It should be obvious that faster rotating background flows will break spherical symmetry and, like with the inclusion of magnetic field, will necessitate a modification to the geometry of the local flow. If, however, such a generalisation could be made to allow for the modelling differentially rotating collapses it would offer the possibility of following a collapse from it's initial nearly spherical cloud all the way to the formation of the disc, with the local model smoothly transitioning from that presented here for the initial collapse to the classic shearing box model at late times which provides a local mode for the newly formed disc.

Finally in the specific application of the local model to protoplanetary disc formation an important additional factor that would be useful to include is dust. If, as expected, the local model proves as as rich as the shearing box model in terms of the number of instabilities and other local phenomena then there should be ample opportunity for local flows with the collapse to concentrate dust - potentially jump starting planet formation by allowing the earliest stages of planet formation to occur prior to the disc being formed, in agreement with the detection of signposts of planet formation in very young stellar systems \citep{Miotello22,Tsukamoto22}. 

\section{Conclusion} \label{conclusion}

In this paper we have developed a local model for a spherically contracting/expanding gas cloud in the same vein as other local models which have achieved widespread usage in astrophysics and planetary science. The model consists of a periodic box with a Cartesian like, but time dependant geometry.

\begin{itemize}
 \item The model is close to the Accelerated Expanding Box model of \citet{Tenerani17}, although we have a different treatment of pressure and density consistent with a small box limit. It is an extension of \citet{Robertson12} to allow for the non-isotropic expansion/contraction expected in spherical flows.
 \item  The model is spatially homogeneous allowing for the use of periodic (or shear-periodic) boundary conditions. 
 \item The model inherits a number of symmetries from fluid dynamics in periodic Cartesian domains and come from the global problem. Notably horizontal Galilean like transforms accelerate with time as consequence of conservation of angular momentum in the global problem.
 \item The energy, density and vorticity of the flow in the box all increase(/decrease) during the collapse(/expansion).
 \item We have derived several nonlinear solutions to our local model. The most important of which are the horizontal shear flow, which correspond to zonal flows in the global problem. These shear flows grow in strength during a collapse as a consequence of conservation of angular momentum.
\end{itemize}

We have suggested several possible extensions to the local model. Our model will hopefully be of some use in the study of the early stages of disc/star formation. However the general utility of the local model will require the development of an effective numerical implementation.

\section*{Acknowledgements}

We would like to thank Beno\^{i}t Commer\c{c}on, Matthew Kunz and Pascal Wang for invaluable help with the literature; Ziyan Xu and Timoth\'{e}e Cl\'{e}ris for discussion pertaining to requirements for numerical implementations and Francesco Lovascio for useful discussions/sanity checking some of the more unexpected properties of the model. We thank the reviewer for there suggestions which helped improve the clarity of this paper.

 E. Lynch would like to thank the European Research Council (ERC). This research was supported by the ERC through the CoG project PODCAST No 864965. This project has received funding from the European Union’s Horizon 2020 research and innovation programme under the Marie Skłodowska-Curie grant agreement No 823823. 

\section*{Data availability}

No new data were generated or analysed in support of this research.




\bibliographystyle{mnras}
\bibliography{PAPER_LOCAL_MODEL_SPHERICALFLOW} 




\onecolumn

\appendix

\section{Reformulated equations for easier numerical implementation} \label{alternative formalisms}

An issue faced by our local model, which it shares with the distorted shearing box models \citep[e.g.][]{Ogilvie13a,Ogilvie14}, is how to include the time dependant geometry in existing solvers. This has hindered the implementation of local models in the past and has led to workarounds such as \citet{Latter06} and \citet{Ogilvie22}. In this appendix we derive various (equivalent) alternative formalism for the local model that may facilitate it's implementation  in existing hydrocodes.

\subsection{Modified continuity equations}

The momentum equation can be written in Cartesian like geometry which removes the time dependant terms in front of the derivatives, by use of an asymmetric stress tensor $T_{i j}$,

\begin{align}
D v^{x} + \frac{2 U_0}{\mathcal{R}} v^{x} &= \frac{1}{\rho} \partial_{x} T^{x x} , \\
D v^{y} + \frac{2 U_0}{\mathcal{R}} v^{y} &= \frac{1}{\rho} \partial_{y} T^{y y} , \\
D v^{\tilde{z}} + 2 v^{\tilde{z}} U_{R 0} &= \frac{1}{\rho} \partial_{\tilde{z}} T^{\tilde{z} \tilde{z}} ,
\end{align}
where 

\begin{equation}
 T^{x x} = T^{y y} = -\frac{p}{\mathcal{R}^2} , \quad T^{\tilde{z} \tilde{z}} = -\frac{p}{L_z^2} , \quad T^{x y} = T^{x \tilde{z}} = T^{y \tilde{z}} = 0 .
\end{equation}
One way to interpret this tensor is as an anisotropic sound speed with sound waves propagating at different speeds in the horizontal and vertical directions. Alternatively we can interpret this as a modified equation of state with stresses from pseudo-self gravity or pseudo-magnetic field based on the following stress tensor for a self-gravitating, ideal MHD fluid,

\begin{equation}
T^{i j} = -p^{\prime} \delta^{i j} - \frac{1}{4 \pi G} \left(g^i g^j - \frac{1}{2} g^2 \delta^{i j} \right) + \frac{1}{\mu_0} \left(B^{i} B^{j} - \frac{1}{2} B^2 \delta^{i j} \right) .
\end{equation}
Substituting the expressions for the stress tensor components into the above expression we see that we require $g^{x} = g^{y} = B^{x} = B^{y} = 0$ in order that the cross terms, $T^{x y}$, $T^{x z}$ and $T^{y z}$ vanish. Adding $T^{x x} + T^{z z}$ (or equivalently $T^{y y} + T^{z z}$) we arrive at an expression relating the old pressure, $p$, to the new pressure, $p^{\prime}$,

\begin{equation}
 p^{\prime} = \frac{1}{2} \left(\frac{1}{\mathcal{R}^2} + \frac{1}{L_z^2} \right) p .
\end{equation}
Subtracting $T^{x x} - T^{z z}$ (equivalently $T^{y y} - T^{z z}$) we obtain the anisotropic part of the stress which is independent of $p^{\prime}$. When $L_z \ge \mathcal{R}$ we can interpret this as a vertical field

\begin{equation}
B^z = \sqrt{\mu_0 p} \left(\frac{1}{\mathcal{R}^2} - \frac{1}{L_z^2} \right)^{1/2} ,
 \label{B_z pseudo field}
\end{equation}
with $g^z = 0$. While if $\mathcal{R} \ge L_z$ we can instead interpret this as a vertical self-gravitational acceleration,

\begin{equation}
g^z = \sqrt{4 \pi G p} \left(\frac{1}{L_z^2}  - \frac{1}{\mathcal{R}^2} \right)^{1/2} ,
\end{equation}
with $B^z = 0$. Note that that neither of these are real magnetic or self-gravity fields as they do not evolve in the correct way, or have the appropriate properties. For example the pseudo-magnetic field given by Equation \ref{B_z pseudo field} is not, typically, divergence free.

For more general spherical flows where these inequalities cannot be guaranteed we can instead include a mixture of effective vertical self-gravitating acceleration of the following form,

\begin{align}
B^z &= \sqrt{\mu_0 p} \mathcal{R}^{-1},\\
g^z &= \sqrt{4 \pi G p} L_{z}^{-1} .
\end{align}

\subsection{Relative density}

For many numerical implementations (particularly Lagrangian/Particle based methods) it is preferable that there is no modification to the continuity equation. One can achieve this, for certain equations of state, by rewriting the continuity equation in terms of the relative density $\psi = \rho/\rho_{\rm ref}$ such that

\begin{equation}
D \psi = - \psi \partial_i v^i .
\end{equation}
The reference density is that of the basic state with $\rho_{\rm ref} = \rho_{\rm ref} (\tau)$, which evolves according to

\begin{equation}
D \rho_{\rm ref} = - \rho_{\rm ref} \Delta .
\end{equation}
Assuming the equation of state is polytropic, and takes the form $p = K(x,y,z) \rho^{1 + 1/n}$, which is appropriate for both adiabatic ($1+1/n = \gamma$) and locally isothermal ($n \rightarrow \infty$) equations of state, then we can rewrite the equation of state in terms of the relative density by introducing the reference pressure $p_{\rm ref} (x,y,z,t) = K(x,y,z) \rho_{\rm ref}^{1 + 1/n}$. The equation of state is then $p = p_{\rm ref} \psi^{1 + 1/n}$. The fluid equations are then

\begin{align}
D v^{x} + \frac{2 U_0}{\mathcal{R}} v^{x} &= - \frac{1}{\psi \mathcal{R}^{2}} \partial_{x} (p/\rho_{\rm ref}) ,\\
D v^{y} + \frac{2 U_0}{\mathcal{R}} v^{y} &= - \frac{1}{\psi \mathcal{R}^{2}} \partial_{y} (p/\rho_{\rm ref})  , \\
D v^{\tilde{z}} + 2 v^{\tilde{z}} U_{R 0} &= - \frac{1}{\psi L_z^2} \partial_{\tilde{z}} (p/\rho_{\rm ref})  , \\ 
D \psi &= - \psi \left[ \partial_x v^{x} + \partial_y v^{y} + \partial_{\tilde{z}} v^{\tilde{z}} \right] .
\end{align}
These equations can be derived from the following Lagrangian

\begin{equation}
 L = \int \psi \left[ \frac{1}{2} \mathcal{R}^2 \left( v^{x} v^{x} + v^{y} v^{y} \right) + \frac{1}{2} L_z v^{\tilde{z}} v^{\tilde{z}} - n \frac{p_{\rm ref}}{\rho_{\rm ref}} \psi^{1/n} \right] \, d x \, d y \, d \tilde{z} ,
\end{equation}
which is useful for deriving smooth particle hydrodynamics base methods. The main difference between this Lagrangian and the more commonly used Lagrangian for ideal fluid is the presence of the time dependant coefficients, $\mathcal{R}$, $L_z$, $p_{\rm ref}$ and $\rho_{\rm ref}$.

\section{Analysis of the Linear Waves} \label{linear wave analysis}

\subsection{Exact solutions for sound waves when decoupled from the vortical waves} \label{beyond WKB}

For many simple spherical flows it is possible to obtain exact solutions for sound waves when they are decoupled from the vortical wave (i.e. with a coupling coefficient $g=0$). This occurs for both purely horizontal and purely vertical waves, along with background flow with a constant aspect ratio (the situation considered in cosmological fluids). To describe the background flow we consider various functional forms for the reference lengthscale $L = L (\tau)$ and obtain solutions for the sound waves in the local model. We adopt

\begin{equation}
 L = L_0 \frac{|\mathbf{k}_0|}{|\mathbf{k}|} = L_0 \frac{\mathcal{R}}{\mathcal{R}_0} \left( \frac{k_x^2 + b_0^{-2} k^2_{\tilde{z}}}{k_x^2 + b^{-2} k^2_{\tilde{z}}}\right)^{1/2} ,
\end{equation}
as our reference lengthscale, where $|\mathbf{k}|$ denotes the magnitude of the wavevector and $|\mathbf{k}_0|$ is the initial $|\mathbf{k}|$. For a horizontally propagating wave choosing $L_0 = R_0$ results in $L = \mathcal{R}$. Similarly for vertically propagating waves, $L_0 = L_{z 0}$ results in $L = L_{z}$. Finally for a constant aspect ratio ($b = b_0$) the reference length can be identified with the scale factor $a$.

We consider two forms of power law profiles, given by

\begin{align}
 L &= L_0 (\tau/\tau_{\rm c})^{\beta} , \label{power law 1} \\
 L &= L_0 (1 - \tau/\tau_{\rm c})^{\beta}  , \label{power law 2}
\end{align}
the latter includes the linear/uniform collapse/expansion when $\beta=1$. On can also consider an exponential collapse/expansion,

\begin{equation}
 L = L_{0} \exp \left(H \tau \right) ,
\label{exponential}
\end{equation}
where $H$ is a constant Hubble parameter. The exponential profile can be obtained from Equation \ref{power law 2} by taking $\tau_c = -H^{-1} \beta$ and taking the limit $\beta \rightarrow \infty$. A logarithmic profile can be similarly obtained by setting $L_0 = A \beta$, for some constant $A$, and taking the limit $\beta \rightarrow 0$. Finally, although we shall not explicitly obtain sound wave solutions for it, the early stages of the isothermal sphere model of \citet{Shu77} leads to the background radius evolving according to

\begin{equation}
 \mathcal{R}^2 = L^2 = L_0^2 \left[1  - (\tau/\tau_c)^2 \right] .
 \label{shu model}
\end{equation}

The dynamics of the sound wave, when it is decoupled from the vortical wave, can be obtained from the following Hamiltonian:

\begin{equation}
 \mathcal{H} = \frac{1}{2} c_s^{-2} \omega^2 (\tau) P_{\alpha}^2 + \frac{1}{2} c_s^2 \alpha^2 ,
\end{equation}
where $\omega (\tau) = c_s |\mathbf{k}_0| L_0/L$ is the sound wave frequency. This equation can be obtained from Equation \ref{sound wae hamiltonian} by setting the coupling term, $c_s^{-1} g(\tau) P_{\alpha} P_{\beta}$, to zero. Hamilton's equations lead to

\begin{align}
 \partial_{\tau} P_{\alpha} &= -  c_s^2 \alpha, \\
\partial_{\tau} \alpha &=  c_s^{-2} \omega^2 P_{\alpha},
\end{align}
from which we obtain the following, parametric oscillator, equation for the momentum:

\begin{equation}
\partial_{\tau}^2 P_{\alpha} + \omega^2 P_{\alpha} = 0 .
\end{equation}
For both the powerlaw profiles the evolutionary equation for the momentum is

\begin{equation}
 \partial_{\tau}^2 P_{\alpha}+ \omega_0^2 s^{-2 \beta/\alpha}  P_{\alpha} = 0 , \label{pix evo}
\end{equation}
where $\omega_0 = c_s |\mathbf{k}| (\tau_0/\tau_c)^{-\beta}$ and we have introduced $s$ which for the Equation \ref{power law 1} is given by $s = (\tau/\tau_0)^{\alpha}$ and for Equation \ref{power law 2} is given by $s = (\frac{\tau_{\rm_c} - \tau}{\tau_{\rm 0}})^{\alpha}$. In both cases this leads to the reference lengthscale taking the form $L = L_0 (\tau_0/\tau_c)^{\beta} s^{\beta/\alpha}$. The characteristic timescale relevant to this, $t_{\rm bg} = L/|\dot{L}|$, is given by

\begin{equation}
t_{\rm bg} = \frac{\tau_0 s^{1/\alpha}}{|\beta|} ,
\end{equation}
by setting $\omega t_{\rm bg} = 1$, and writing in terms of $L$, we can calculate the reference lengthscale at freeze out (provided $\beta \ne 1$),

\begin{equation}
 L_{\rm freeze out} = L_0 \left(\frac{\tau_0}{\tau_c}\right)^{\beta} \left( \frac{|\beta|}{\omega_0 \tau_0} \right)^{\beta/(1 - \beta)} ,
\end{equation}
which gives the value of $L$ at which a mode of a given wavelength enters (or exits) the freeze out regime.

Equation \ref{pix evo} can be transformed into Bessel's equation by setting $P_{\alpha} = A s^{\lambda} f (s)$, resulting in the following equation for $f$,

\begin{equation}
s^{2} f^{\prime \prime} (s)  + \left [ 1 - 1/\alpha +  2 \lambda \right] s f^{\prime} (\tau) +   \left[  \lambda (\lambda - 1/\alpha) + \omega_0^2 \alpha^{-2} \tau_0^{2} s^{2(1 - \beta)/\alpha}  \right] f (\tau)  = 0 ,
 \label{transformed equation}
\end{equation}
where we have made use of $\ddot{s} s \dot{s}^{-2} = 1 - 1/\alpha$  and $\dot{s}^{2} =  \alpha^{2} \tau_0^{-2} s^{2 - 2/\alpha}$.

For $\beta \ne 1$, Equation \ref{transformed equation} corresponds to Bessel's equation provided that $\alpha = 1 - \beta$, $\lambda  = \frac{1}{2 (1 - \beta)}$ and 

\begin{equation}
 \tau_0 = \left[ \frac{|1 - \beta|}{ c_s |\mathbf{k}_0| \tau_c^{\beta} } \right]^{1/(1-\beta)} ,
\end{equation}
This results in the following general solution for $P_{\alpha}$,

\begin{equation}
 P_{\alpha} = s^{1/[2( 1 - \beta)]} \left[ A  J_{\nu} (s) + B  Y_{\nu} (s) \right] ,
\end{equation}
where $J_{\nu}$ and $Y_{\nu}$ are Bessel functions with order $\nu = \frac{1}{2 |1 - \beta|}$. The solution for the exponential profile can be obtained from the above by taking the limit $\beta \rightarrow \infty$. This can, alternatively, be written in terms of the reference length, 

\begin{equation}
 P_{\alpha} =  L^{1/(2\beta)}\left\{ \tilde{A} J_{\nu} \left[ \left| \frac{1 - \beta}{\beta}\right| \left( \frac{L}{L_{\rm freeze out} }\right)^{(1-\beta)/\beta} \right] + \tilde{B}  Y_{\nu} \left[ \left| \frac{1 - \beta}{\beta}\right| \left( \frac{L}{L_{\rm freeze out} }\right)^{(1-\beta)/\beta} \right]   \right\}  ,
\end{equation}
where the freeze out radius can be simplified to

\begin{equation}
 L_{\rm freeze out} = L_0 \left(\frac{\tau_0}{\tau_c}\right)^{\beta} \left| \frac{\beta}{1 - \beta} \right|^{\beta/(1 - \beta)} ,
\end{equation}
and we have absorbed constant terms into $\tilde{A}$ and $\tilde{B}$. In terms of the $s$ variable the velocity and density perturbations, due to the sound wave, are

\begin{align}
 \delta v^{x} &= \frac{2 k_x}{\mathcal{R}^{2} }  s^{1/[2 (1 - \beta)]} \left \{ \mathrm{Re} [A \exp (i \mathbf{k} \cdot \mathbf{x})] J_{\nu} \left( s \right) + \mathrm{Re} [B \exp (i \mathbf{k} \cdot \mathbf{x})] Y_{\nu} \left( s \right)   \right\} , \\
 \delta v^{\tilde{z}} &= \frac{2 k_{\tilde{z}} }{L_{z}^{2}} s^{1/[2 (1 - \beta)]} \left \{ \mathrm{Re} [A \exp (i \mathbf{k} \cdot \mathbf{x})] J_{\nu} \left( s \right) + \mathrm{Re} [B \exp (i \mathbf{k} \cdot \mathbf{x})] Y_{\nu} \left( s \right)   \right\} , \\
 \delta \rho &= - \frac{2 \rho}{c_s^2} \frac{1 - \beta}{\tau_0} s^{\frac{1 - 2 \beta}{2 (1 - \beta)}} \sgn (\dot{s})\left \{ \mathrm{Im} [A \exp (i \mathbf{k} \cdot \mathbf{x})] J_{\nu - 1} \left( s \right) + \mathrm{Im} [B \exp (i \mathbf{k} \cdot \mathbf{x})] Y_{\nu - 1} \left( s \right)   \right\} .
\end{align}
These solutions are very similar to linear perturbation to FRW-cosmology in the presence of a perfect fluid (e.g. see \citet{Baumann12}). This similarity is mostly a reflection of the similar mathematical structure (2nd-order linear differential equations with powerlaw coefficients). In both cases the background is a spatially homogeneous perfect fluid in an FRW-(like) metric. In our local model we consider perturbations to the fluid holding the metric fixed. In cosmological perturbation theory, however, one is instead considering perturbations to the metric in the presence of the fluid.

The above solutions work for $\beta \ne 1$. For $\beta = 1$ the above solution is singular as $\nu \rightarrow \infty$. Without loss of generality we can set $\alpha = 1$ and one sets $\lambda = 0$ leading to

\begin{equation}
 s^2 f^{\prime \prime} + \left( c_s |\mathbf{k}_0| \tau_c \right)^{2} f = 0 .
\end{equation}
This can be solved by setting $f = s^{\gamma}$, provided

\begin{equation}
 \gamma (\gamma - 1) = - \left( c_s |\mathbf{k}_0| \tau_c \right)^{2} ,
\end{equation}
resulting in the general solution

\begin{equation}
 P_{\alpha} = \sqrt{s} \left[ A_{+} s^{\lambda} + A_{-} s^{-\lambda }  \right] ,
\end{equation}
where we have introduced $\lambda = \frac{1}{2} \sqrt{1 - 4 \left( c_s |\mathbf{k}_0| \tau_c \right)^{2}}$ In terms of the reference lengthscale the solution is
\begin{equation}
 P_{\alpha} = \sqrt{L} \left[ \tilde{A}_{+} L^{\lambda } + \tilde{A}_{-} L^{-\lambda }  \right] ,
\end{equation}
where, again, we have absorbed constant terms into $\tilde{A}_{+}$ and $\tilde{A}_{-}$. From this expression we can obtain the following density and velocity perturbations

\begin{align}
\delta v^{x} &= \frac{2}{\mathcal{R}^2} k_x \sqrt{L} \left \{ \mathrm{Re} \left[ \tilde{A}_{+} L^{\lambda} \exp \left( i \mathbf{k} \cdot \mathbf{x} \right) \right ] + \mathrm{Re} \left[ \tilde{A}_{-}  L^{-\lambda}  \exp \left( i \mathbf{k} \cdot \mathbf{x}\right) \right ] \right\} , \\
 \delta v^{\tilde{z}} &= \frac{2}{L_{z}^2} k_{\tilde{z}} \sqrt{L} \left \{ \mathrm{Re} \left[ \tilde{A}_{+}  L^{\lambda} \exp \left( i \mathbf{k} \cdot \mathbf{x} \right) \right ] + \mathrm{Re} \left[ \tilde{A}_{-}  L^{-\lambda}  \exp \left( i \mathbf{k} \cdot \mathbf{x}\right) \right ] \right\} , \\
\delta \rho &= \frac{2 \rho L_0}{c_s^2 \tau_c} L^{-1/2} \left \{ \mathrm{Im} \left[ \left( \lambda + 1/2 \right) \tilde{A}_{+}  L^{\lambda} \exp \left( i \mathbf{k} \cdot \mathbf{x} \right) \right ] - \mathrm{Im} \left[ \left( \lambda - 1/2 \right) \tilde{A}_{-}  L^{-\lambda}  \exp \left( i \mathbf{k} \cdot \mathbf{x} \right) \right ] \right\} .
\end{align}
This solution is useful to illustrate the effects of freeze-out on the wave. For large $|\mathbf{k}_0|$, $\lambda$ is imaginary and the $+$ and $-$ solutions corresponds to left and right travelling waves which grow during a collapse. Reducing $|\mathbf{k}_0|$ induces a phase shift between the density and velocity perturbation. When $|\mathbf{k}_0|$ drops below $k_{\rm freezeout} = \frac{1}{2 c_s \tau_c}$ then the wave enters the freeze-out regime where the left and right travelling waves become standing waves with differing growth(/decay)-rates.

Solutions for horizontally propagating sound waves in the isothermal sphere collapse of \citet{Shu77} (Equation \ref{shu model}) can similarly be found in terms of Hypergeomtric functions, however we shall not do this here.

\subsection{Diagonally propagating waves and sound wave - vortical wave coupling} \label{diagonally propergating waves}

The density and velocity perturbations with wavenumber $\pm\mathbf{k}$ can be obtained from the $(\alpha, \beta, \gamma,  P_{\alpha},P_{\beta}, P_{\gamma})$ system of coordinates and momenta as follows

\begin{align}
 \delta v^{x} &= \frac{2}{\mathcal{R}^2} \, \left \{ k_x \mathrm{Re} \left[ P_{\alpha} \exp (i \mathbf{k} \cdot \mathbf{x} )\right] + \frac{k_{\tilde{z}}}{\sqrt{k_x^2 b_0^2 + k^2_{\tilde{z}}}} \mathrm{Re} \left[ P_{\beta} \exp (i \mathbf{k} \cdot \mathbf{x} )\right] \right\} , \\
 \delta v^{y} &= \frac{2}{\mathcal{R}^2} \, \mathrm{Re} \left[ P_{\gamma} \exp (i \mathbf{k} \cdot \mathbf{x} )\right]  \\
 \delta v^{\tilde{z}} &=  \frac{2}{L_z^2} \, \left \{ k_{\tilde{z}} \mathrm{Re} \left[ P_{\alpha} \exp (i \mathbf{k} \cdot \mathbf{x} )\right] - \frac{k_{x} b_0^2}{\sqrt{k_x^2 b_0^2 + k^2_{\tilde{z}}}} \mathrm{Re} \left[ P_{\beta} \exp (i \mathbf{k} \cdot \mathbf{x} )\right] \right\} , \\
 \delta \rho &= 2 \rho \, \mathrm{Im} \left [ \alpha \exp \left( i \mathbf{k} \cdot \mathbf{x} \right) \right] ,
\end{align}
where we have made use of the reality of the perturbations where we require $\delta \mathbf{v}^{*} = \delta \mathbf{v}$ and $\delta \rho^{*} = \delta \rho$. $P_{\beta}$ and $P_{\gamma}$ are integrals of motion related to the fluid vorticity. The dynamics of the $(\alpha,P_{\alpha})$are obtained from the following Hamiltonian,

\begin{equation}
 \mathcal{H} = \frac{1}{2} c_s^{-2} \omega^2 (\tau) P_{\alpha}^2 + c_s^{-1} g (\tau) P_{\alpha} P_{\beta} + \frac{1}{2} c_s^2 \alpha^2 ,
\end{equation}
where the sound wave frequency, $\omega$, and coupling coefficient, $g$, are given by Equation \ref{sound wave freq} and \ref{coupling coefficient} respectively. 

Neglecting $P_{\gamma}$ we have the following WKB solutions for $\delta \rho$ and $\delta \mathbf{v}$,

\begin{align}
 \delta v^{x} &\approx - \frac{2 c_s}{\mathcal{R}^2} \omega^{-1/2} k_x \mathrm{Re} \left[ X_{+} \exp \left(i \mathbf{k} \cdot \mathbf{x} + i\int \omega d \tau \right) + X_{-} \exp \left( i \mathbf{k} \cdot \mathbf{x} - i\int \omega d \tau \right)\right] + \frac{2}{\mathcal{R}^2} \frac{k_{\tilde{z}}}{\sqrt{k_x^2 b_0^2 + k^2_{\tilde{z}}}} \mathrm{Re} \left[ P_{\beta} \exp \left( i \mathbf{k} \cdot \mathbf{x}  \right) \right] \\
 \delta v^{\tilde{z}} &\approx - \frac{2 c_s}{L_z^2} \omega^{-1/2} k_z \mathrm{Re} \left[ X_{+} \exp \left(i \mathbf{k} \cdot \mathbf{x} + i\int \omega d \tau \right) + X_{-} \exp \left( i \mathbf{k} \cdot \mathbf{x} - i\int \omega d \tau \right)\right]  - \frac{2}{L_z^2} \frac{k_{x} b_0^2}{\sqrt{k_x^2 b_0^2 + k^2_{\tilde{z}}}} \mathrm{Re} \left[ P_{\beta} \exp \left( i \mathbf{k} \cdot \mathbf{x}  \right) \right] \\
 \delta \rho &\approx \frac{2 \rho}{c_s} \omega^{1/2} \mathrm{Re} \left[ X_{+} \exp \left(i \mathbf{k} \cdot \mathbf{x} + i \int \omega d \tau \right) - X_{-} \exp \left( i \mathbf{k} \cdot \mathbf{x} - i\int \omega d \tau \right)\right] ,
\end{align}
with $\delta v^{y} = 0$ and $X_{\pm}$ and $P_{\beta}$ are (complex) constants. For diagonally propagating waves we require $|P_{\beta}| \ll c_s \omega^{-1/2} |X_{\pm}|$ in order for the above approximation to be valid.

It is useful to consider the shear and velocity divergence as a result of the linear waves. The divergence of the velocity is given by

\begin{equation}
\partial_{i} \delta v^{i} = \frac{2}{\mathcal{R}^2} \left( k_x^2 + b^{-2} k_{\tilde{z}}^2 \right) \mathrm{Re} [i P_{\alpha} \exp \left( i \mathbf{k} \cdot \mathbf{x} \right)] + \frac{2}{\mathcal{R}^2} \frac{k_x k_{\tilde{z}}}{\sqrt{ k_x^2 b_0^2 + k^2_{\tilde{z}} }} \left[1 - (b/b_0)^{-2}\right] \mathrm{Re} [i P_{\beta} \exp \left( i \mathbf{k} \cdot \mathbf{x} \right) ] ,
\end{equation}
while the velocity shear (neglecting $\delta v^{y}$) can be characterised by

\begin{equation}
 k_{\tilde{z}} \delta v^{x} - k_{x} \delta v^{\tilde{z}} = \frac{2}{\mathcal{R}^2} \left \{ k_x k_{\tilde{z}} (1 - b^{-2}) \mathrm{Re} [P_{\alpha} \exp(i \mathbf{k} \cdot \mathbf{x})] + \frac{k_x^2 b^2 + k^2_{\tilde{z}}}{\sqrt{k_x^2 b_0^2 + k^2_{\tilde{z}}}} \mathrm{Re} [P_{\beta} \exp(i \mathbf{k} \cdot \mathbf{x})]  \right\} . \label{shear measure}
\end{equation}
Diagonal vortical waves, with $P_{\alpha} (0) = \alpha (0) = 0$, cause a divergence of the velocity (and thus a density perturbation) when the coupling coefficient $g \ne 0$. This occurs as a result of a varying box aspect ratio. Similarly, from Equation \ref{shear measure}, we see that diagonal sound waves (with $P_{\beta} = 0$) generate a shear flow perpendicular to the direction of propagation if the aspect ratio $b \ne 1$. For a constant aspect ratio ($b = b_0$) this is simply a consequence of the choice of coordinate system and doesn't reflect a physical shear in the fluid flow. However when the aspect ratio is time dependant this shear cannot be removed by a suitable rescaling of the coordinates, meaning the propagating sound waves can generate velocity shear in flows with a time varying $b$.

\textbf{\section{Local models of some analytical background flows} }

In this section we shall consider some example spherical flows and show how to obtain the geometrical coefficients necessary to formulate the local model. A wide variety of spherical problems can be described by powerlaw flows of the form

\begin{equation}
 U = \mathcal{U}_0 (R/R_0)^{\beta} ,
\end{equation}
where $\mathcal{U}_0$, $R_0$ and $\beta$ are constants, over some range of radii. This includes the free-fall case $\beta=-1/2$, the \citet{Penston69} solution away from the transition point ($\beta=1/7$ at large $R$ and $\beta=1/2$ at small $R$) and the solution of \citet{Hernandez22} ($\beta=-1/2$). Steady flows require $\rho_0 \propto R^{-(2 + \beta)}$, however the local model is agnostic to whether the flow is a steady state.

We shall also consider the isothermal spherical collapse problem due to \citet{Shu77}. This has a radial velocity at large radii/early times of

\begin{equation}
 U = - a^2 (A - 2) t R^{-1} .
\end{equation}
where $a$ and $A$ have the same meaning as in \citet{Shu77}.

Starting with the powerlaw flow. The reference radius can be obtained from

\begin{equation}
 \dot{\mathcal{R}} = U(\mathcal{R}) = \mathcal{U}_0 R_0^{-\beta} \mathcal{R}^{\beta} ,
\end{equation}
which has solution

\begin{equation}
 \mathcal{R} = R_0 \left[1  + (1 - \beta) \mathcal{U}_0 R_0^{-1} t \right]^{1/(1 - \beta)} .
\end{equation}
This allows us to determine $U_0$ and $U_{R 0}$,

\begin{equation}
 U_0 = U(\mathcal{R}) =  \mathcal{U}_0 \left[1  + (1 - \beta) \mathcal{U}_0 R_0^{-1} t \right]^{\beta/(1 - \beta)} ,
\end{equation}

\begin{equation}
 U_{R 0} = \frac{\partial U}{\partial R} \Biggr|_{R = \mathcal{R}} = \beta \mathcal{U}_0 R_0^{-1} \left[1  + (1 - \beta) \mathcal{U}_0 R_0^{-1} t \right]^{ -1 }  .
\end{equation}
Substituting the expression for $U_{R 0}$ into the evolutionary equation for $L_z$, $\dot{L}_z = U_{R 0} L_z$, we obtain an expression for $L_z$:

\begin{equation}
 L_z = L_{z 0} \left[1  + (1 - \beta) \mathcal{U}_0 R_0^{-1} t \right]^{\beta/(1 - \beta)} ,
\end{equation}
with $L_{z 0}$ the initial vertical lengthscale. The divergence of the background flow, $\Delta$, is then

\begin{equation}
 \Delta = (2 + \beta) \mathcal{U}_0 R_0^{-1} \left[1  + (1 - \beta) \mathcal{U}_0 R_0^{-1} t \right]^{ -1 }  .
\end{equation}
The basic state for this model is $v = 0$ and a spatially homogeneous density given by

\begin{equation}
 \rho = \rho_{\rm init} \left[1  + (1 - \beta) \mathcal{U}_0 R_0^{-1} t \right]^{ -(2 + \beta)/(1 - \beta) } ,
\end{equation}
where $\rho_{\rm init}$ is the initial density.

For the model of \citet{Shu77} we can similarly obtain an expression for $\mathcal{R}$,

\begin{equation}
 \mathcal{R} = R_0 \left[ 1 - \frac{a^2}{R_0^2} (A - 2) t^2 \right]^{1/2} .
\end{equation}
We thus obtain expressions for $U_0$ and $U_{R 0}$,

\begin{equation}
 U_0 = - a^2 (A - 2) R_0^{-1} t \left[ 1 - \frac{a^2}{R_0^2} (A - 2) t^2 \right]^{-1/2} , \quad U_{R 0} = \frac{a^2}{R_0^2} (A - 2) t \left[ 1 - \frac{a^2}{R_0^2} (A - 2) t^2 \right]^{-1} .
\end{equation}
From these we obtain expressions for $L_z$,

\begin{equation}
 L_z = L_{z 0} \left[ 1 - \frac{a^2}{R_0^2} (A - 2) t^2 \right]^{-1/2} ,
\end{equation}
while the divergence of the background flow is

\begin{equation}
 \Delta = -\frac{a^2}{R_0^2} (A - 2) t \left[ 1 - \frac{a^2}{R_0^2} (A - 2) t^2 \right]^{-1} .
\end{equation}
The basic state of the local model for the \citet{Shu77} isothermal collapse is $v=0$ and a spatially homogeneous density given by

\begin{equation}
 \rho = \rho_{\rm init}  \left[ 1 - \frac{a^2}{R_0^2} (A - 2) t^2 \right]^{-1/2} .
\end{equation}

This appendix has shown how to go from fairly simple collapse profiles (powerlaw and the isothermal sphere model)  to the geometrical terms necessary to formulate the local model. More general global collapse solutions can only be obtained numerically meaning there will be no closed form solution for the geometric terms. However, the principles of obtaining the geometric terms remain the same and can be expressed as numerical solutions to the ODEs $\dot{\mathcal{R}} = U_0(\mathcal{R}, t)$ and $\dot{L}_z = U_{R 0} L_z$.


\bsp	
\label{lastpage}
\end{document}